\documentclass[aps,twocolumn,prb,amsmath,amssymb,superscriptaddress]{revtex4-1}
\usepackage{url}
\usepackage{graphicx}
\usepackage{color}
\usepackage{natbib}
\usepackage{hyperref}
\usepackage{notes2bib}
\usepackage{mathbbol}
\usepackage{float}
\usepackage{ulem}
\usepackage{tikz}
\usepackage{xcolor}
\usepackage{svg}
\usepackage{ulem}

\begin{document}

\title{Superconductivity in monolayer and few-layer graphene: \\ II. Topological edge states and Chern numbers}

\author{Adeline Cr\'epieux}
\affiliation{Aix Marseille Univ, Universit\'e de Toulon, CNRS, CPT, Marseille, France}
\author{Emile Pangburn}
\affiliation{Institut de Physique Th\'eorique, Universit\'e Paris Saclay, CEA
CNRS, Orme des Merisiers, 91190 Gif-sur-Yvette Cedex, France}
\author{Louis Haurie}
\affiliation{Institut de Physique Th\'eorique, Universit\'e Paris Saclay, CEA
CNRS, Orme des Merisiers, 91190 Gif-sur-Yvette Cedex, France}
\author{Oladunjoye A.~Awoga}
\affiliation{Solid State Physics and NanoLund, Lund University, Box 118, S-221 00 Lund, Sweden}
\author{Annica M. Black-Schaffer}
\affiliation{Department of Physics and Astronomy, Uppsala University, Box 516, S-751 20 Uppsala, Sweden}
\author{Nicholas Sedlmayr}
\affiliation{Institute of Physics, Maria Curie-Sk\l{}odowska University,
Plac Marii Sk\l{}odowskiej-Curie 1, PL-20031 Lublin, Poland}
\author{Catherine P\'epin}
\affiliation{Institut de Physique Th\'eorique, Universit\'e Paris Saclay, CEA
CNRS, Orme des Merisiers, 91190 Gif-sur-Yvette Cedex, France}
\author{Cristina Bena}
\affiliation{Institut de Physique Th\'eorique, Universit\'e Paris Saclay, CEA
CNRS, Orme des Merisiers, 91190 Gif-sur-Yvette Cedex, France}

\date{\today}

\begin{abstract}
We study the emergence of electronic edge states in superconducting (SC) monolayer, bilayer, and trilayer graphene for both spin-singlet and spin-triplet SC order parameters. 
We focus mostly on the gapped chiral $p+ip'$- and $d+id'$-wave SC states that show a non-zero Chern number and a corresponding number of edge states. For the $p+ip'$-wave state, we observe a rich Chern phase diagram when tuning the chemical potential and the SC order parameter amplitudes, which depends strongly on the number of layers and their stacking, and is also modified by trigonal warping. At small parameter values we observe a region whose Chern number is unique to rhombohedrally stacked graphene, and is independent of the number of layers. 
Our results can be understood in relation not only to the SC order parameter winding as expected, but also to the normal state band structure.  This observation establishes the importance of the normal state characteristics for understanding the topology in SC graphene systems.
\end{abstract}

\maketitle

%%%%%%%%%%%%%%%%%%%%%%%%%%%%%%%%%%%%%%%%%%%%%%%%%%%%%%%%%%%%%%%%%%%%%%%%%%%%%%%%%%%%%%%%%%%%%%%%%%%%%%%%%%%%
%                                                                                                          %
%                                                                                                          %
%                           INTRODUCTION                                                    %
%                                                                                                          %
%                                                                                                          %
%%%%%%%%%%%%%%%%%%%%%%%%%%%%%%%%%%%%%%%%%%%%%%%%%%%%%%%%%%%%%%%%%%%%%%%%%%%%%%%%%%%%%%%%%%%%%%%%%%%%%%%%%%%%

\section{Introduction}

Recently there has been a lot of interest in the study of apparently exotic superconductivity (SC) arising in twisted bilayer graphene\cite{Wong2021,cao2018unconventional,po2018origin,lu2019superconductors,balents2020superconductivity,andrei2020graphene,oh2021evidence,cao2021nematicity,christos2020superconductivity,chichinadze2020nematic,wu2020harmonic,yu2021nematicity,fischer2022unconventional}, as well as in rhombohedral multilayer graphene\cite{Zhou2021, ghazaryan2021unconventional}. The apparent similarities to the physics of the cuprate superconductors make one wonder if the study of SC in graphene may shed light also on the physics of exotic high-temperature superconductors and eventually help with creating superconductors with higher and higher critical temperatures. One of the main pressing questions to be answered in this regard is the nature of the SC state in different graphene systems. Many different proposals have been made, for both spin-singlet and spin-triplet order parameters, beyond conventional $s$-pairing, such as $d$- and $p$-wave states\cite{BlackSchaffer2007, Nandkishore2012, Kiesel2012, Chubukov2020, Ulloa2022, Sachdev2020,alidoust2019symmetry,alidoust2020josephson,thingstad2020phonon,milanovici}. Understanding the type of SC order parameter developed by these systems may help in making progress in understanding the underlying SC mechanism, be it phonon-based and enhanced by the formation of flat bands in the multilayer graphene spectrum, such as in twisted bilayer \cite{Bernevig2019} or rhombohedral graphene\cite{Sarma2021}, or electron-electron interaction induced and also enhanced by the flat band formation \cite{kennes2018strong}. Depending on the nature of the SC order parameter, the question also arises about the possibility of generating topological edge and corner states, and eventually putative Majorana states.

In a recent work \cite{Pangburn2022} we have examined the most energetically favorable symmetry-allowed SC states in monolayer graphene, as well as in bilayer and ABA and ABC trilayer graphene, in order to lay the basis for the understanding of their fundamental characteristics. Thus we have calculated the band structure which would be visible using angle-resolved photoemission spectroscopy (ARPES) and the density of states, measurable in scanning tunneling spectroscopy (STS), and in this way distinguished between nodal ($d_{xy}$-, $d_{x^2-y^2}$-, $p_x$-, $p_y$-wave) and gapped ($s_{\rm on}$-, $s_{\rm ext}$-, $p+ip'$-, $d+ id$-, $f$-wave) order parameters, arising from onsite, nearest-neighbor (NN), or next-to-nearest neighbor (NNN) pairing ranges. We have also examined the gap-closing points arising in the spectrum when varying the chemical potential and the SC order parameter. For the multilayer systems we additionally have examined the effects of the trigonal warping. 

In the present work we continue this analysis by studying the non-trivial topology of several of the SC states. We do so by both calculating the Chern number and studying their topologically protected edge states. The main systems we are here focusing on are those exhibiting $d+id'$- and $p+ip'$-wave symmetry order parameters, which are both topological as they are both chiral states \cite{Volovikbook, ReadGreen2000, BlackSchaffer2014} and display a hard energy gap in the spectrum, for which the formation of topological edge states is easy to distinguish. We find that for the $d+id'$-wave SC state, the Chern number is roughly independent of the system parameters such as the chemical potential and the SC order parameter, while for the $p+ip'$-wave state we obtain rich phase diagrams which depend on the number of layers, as well as on the type of stacking for the multilayer systems.  We also find that the trigonal warping has a significant effect on the topological phase diagram at large values of the chemical potential and the SC order parameter. One of our most intriguing observations is that for both $d+id'$ and $p+ip'$ multilayer samples at small chemical potential and SC order parameter there is a small region whose size depends on the value of the interlayer hopping, and whose Chern number for rhombohedral, ABC-stacked, graphene does in fact not depend on the number of layers.  This observation can be used to identify rhombohedral ABC-stacked graphene from other types of stacking.

Our results indicate that the value of the Chern number is not only tied to the winding of the SC order parameter as expected \cite{Volovikbook,ReadGreen2000,BlackSchaffer2014}, but is also strongly correlated to the normal-state band structure. In fact, for a $p+ip'$ order parameter, we note that for chemical potentials below the van Hove singularity, each spin state filled band contributes a Chern number of $2$, since the SC order parameter winds once around each of the two Fermi surfaces localized around the Dirac $K$ and $K^\prime$-points in the Brillouin zone.
%. This is true for both the $p+ip'$- and $d+id'$-wave states. 
On the other hand, for a chemical potential above the van Hove singularity, each spin-state band contributes a Chern number of $1$,
%for $p+ip'$-wave, but $2$ for $d+id'$-wave, 
given that in this regime the system has only one Fermi surface centered around the $\Gamma$ point. For multilayer systems we can likewise predict the Chern number by counting the bands and examining the normal-state Fermi surface topology.

We complement our Chern number calculations by a calculations of the edge states in both chiral SC states. We use a technique developed recently \cite{Bena2019}, which relies on the fact that the spectral function in a semi-infinite system can be obtained by introducing a scalar line-like impurity in an infinite system and using the T-matrix formalism. Our obtained edge energy spectra confirm that the number of observed edge states corresponds to the bulk Chern number following the bulk-boundary correspondence \cite{HasanRMP, QiRMP}. The only exception are a few irregular cases for which extra unprotected zero-energy crossings are observed in the spectrum. Besides the gapped $d+id'$- and $p+ip'$-wave states, we find that the gapless $d_{xy}$-wave system also shows interesting edge properties, related to its nodal superconductivity, which we briefly review.

The rest of the article is organized as follows. In Sec.~II we present the formalism used to describe the SC states in graphene and how to calculate the Chern number  and the correction to the spectral function in the presence of an edge. The results obtained for monolayer, bilayer, and trilayer graphene are discussed in Secs.~III, IV and~V, respectively. We summarize our results in Sec.~VI.

%%%%%%%%%%%%%%%%%%%%%%%%%%%%%%%%%%%%%%%%%%%%%%%%%%%%%%%%%%%%%%%%%%%%%%%%%%%%%%%%%%%%%%%%%%%%%%%%%%%%%%%%%%%%
%                                                                                                          %
%                                                                                                          %
%                           THEORY                                                                %
%                                                                                                          %
%                                                                                                          %
%%%%%%%%%%%%%%%%%%%%%%%%%%%%%%%%%%%%%%%%%%%%%%%%%%%%%%%%%%%%%%%%%%%%%%%%%%%%%%%%%%%%%%%%%%%%%%%%%%%%%%%%%%%%

\section{Model and method}
We start by introducing the model and Hamiltonian for superconductivity in graphene systems. We then introduce the formula for how to calculate the Chern number, followed by the technique of extracting the edge spectral function.

\subsection{Superconducting Hamiltonian}
We are primarily interested in describing the chiral SC states in graphene systems, which can be modeled as pairing on intralayer NN bonds \cite{BlackSchaffer2007,Pangburn2022}.
For single layer graphene with SC pairing induced on NN bonds, the Hamiltonian $H_{\bf k}$ is given by a Fourier transform of the real-space Hamiltonian $H=H_0+H_\text{NN}$, where
\begin{eqnarray}\label{H_0}
\text{H}_{0}&=&-t \sum \limits_{\langle i,j \rangle ,\sigma} [ a^\dagger_{i\sigma} b_{j\sigma} + b^\dagger_{j\sigma} a_{i\sigma}]\nonumber\\
&&- \mu \sum \limits_{i\sigma} \left[a^{\dagger}_{i \sigma}a_{i\sigma}+b^{\dagger}_{i\sigma}b_{i\sigma}\right]~,
\end{eqnarray} 
is the non-interacting Hamiltonian, with $t$ the NN hopping parameter between neighboring A and B sites indexed by $i$ and $j$, $\mu$ the chemical
potential, while $a^\dagger_{i\sigma}$ $(a_{i\sigma})$ and $b^\dagger_{i\sigma}$ $(b_{i\sigma})$ are the creation (annihilation) operators for the A and B lattice sites of the graphene honeycomb lattice. The SC Hamiltonian for the spin-singlet channel ($\eta=0$) and for the spin-triplet channels ($\eta=x,y,z$) is given by\cite{Pangburn2022}
\begin{eqnarray}
H^0_\text{NN}&=&\sum\limits_{\langle ij \rangle}\Delta_{ij}^{0}(a^\dagger_{i\uparrow}b^\dagger_{j\downarrow}-a^\dagger_{i\downarrow}b^\dagger_{j\uparrow})+h.c.~,\\
H^x_\text{NN}&=&\sum \limits_{\langle i,j \rangle}  \Delta_{ij}^{x} (a^\dagger_{i\uparrow} b^\dagger_{j\uparrow} -  a^\dagger_{i\downarrow} b^\dagger_{j\downarrow}) + h.c. ~,\\
H^y_\text{NN}&=&\sum \limits_{\langle i,j \rangle}  \Delta_{ij}^{y} (a^\dagger_{i\uparrow} b^\dagger_{j\uparrow} + a^\dagger_{i\downarrow} b^\dagger_{j\downarrow} )+ h.c.~, \\\label{H_NN}
H^z_\text{NN}&=& \sum \limits_{\langle i,j \rangle} \Delta^{z}_{ij} (a^\dagger_{i\uparrow} b^\dagger_{j\downarrow}  + a^\dagger_{i\downarrow}b^\dagger_{j\uparrow}) + h.c.~,
\end{eqnarray}
where $\Delta_{ij}^{\eta}$ is the NN SC order parameter in channel $\eta$. Fourier transforming into reciprocal space we arrive at
\begin{eqnarray}
H^0_\text{NN}&=& \sum \limits_{k}h_\text{NN}^{0}(\mathbf{k}) ( a^\dagger_{k\uparrow} b^\dagger_{-k\downarrow}  - a^\dagger_{k\downarrow}b^\dagger_{-k\uparrow}) + h.c.~,\\
H^x_\text{NN}&=&\sum \limits_{k} h_\text{NN}^x(\mathbf{k}) (a^\dagger_{k\uparrow} b^\dagger_{-k\uparrow} -  a^\dagger_{k\downarrow} b^\dagger_{-k\downarrow} )+ h.c. ~,\\
H^y_\text{NN}&=& i\sum \limits_{k}h_\text{NN}^y(\mathbf{k})( a^\dagger_{k\uparrow} b^\dagger_{-k,\uparrow}+ a^\dagger_{k,\downarrow} b^\dagger_{-k\downarrow}) + h.c. ~, \\
H^z_\text{NN}&=& \sum \limits_{k}h_\text{NN}^z(\mathbf{k})(a^\dagger_{k\uparrow}b^\dagger_{-k\downarrow}+ a^\dagger_{k\downarrow}b^\dagger_{-k\uparrow})+ h.c. ~,
\end{eqnarray}
where the form factors $h_\text{NN}^{\eta}(\mathbf{k})$ are given in Table~\ref{table1} for the lowest harmonic of all symmetry-allowed states.
Setting $t = 1$ as the energy scale, and the distance between two neighboring carbon atoms, $a=1$, as the length scale, the only remaining tunable parameters are the chemical potential $\mu$ and the SC order parameter amplitude $\Delta_0$.

\begin{table}
\renewcommand*\arraystretch{1.8}
\resizebox{8.7cm}{!}{
\begin{tabular}{|c|l|l|}
  \hline
  $\quad\eta\quad$ & Symmetry & Form factor $h_\text{NN}^{\eta}(\mathbf{k})$ \\
  \hline
 \;$0$ & \;$s_\text{ext}$ & $ h_\text{NN}^{0,s_\text{ext}}({\mathbf{k}})= \frac{\Delta_0}{\sqrt{3}} \tilde{h}_0({\mathbf{k}})$  \\
 \;$0$ & \; $d_{x^2-y^2}$ & $ h_\text{NN}^{0,d_{x^2-y^2}}({\mathbf{k}})= \frac{2\Delta_0}{\sqrt{6}}e^{-ik_y}\left[1-e^{\frac{3i}{2}k_y}\cos(\frac{\sqrt{3}}{2}k_x)\right]$ \\
\;$0$ & \; $d_{xy}$ & $ h_\text{NN}^{0,d_{xy}}({\mathbf{k}})=  \Delta_{0}\sqrt{2}  i \ e^{\frac{i}{2}k_y}\sin(\frac{\sqrt{3}}{2}k_x)$ \\ 
  \hline
 \; $x$ & \;$p_y$ & $h_\text{NN}^{\eta,p_y}({\mathbf{k}})= \frac{2 \Delta_0}{\sqrt{6}}e^{-ik_y}\left[1-e^{\frac{3i}{2}k_y}\cos(\frac{\sqrt{3}}{2}k_x)\right]$  \\
 \; $x$ & \; $p_x$&  $h_\text{NN}^{\eta,p_x}({\mathbf{k}})= i\sqrt{2}\Delta_0 e^{\frac{i}{2}k_y}\sin(\frac{\sqrt{3}}{2}k_x)$ \\
  \hline
\end{tabular}}\caption{Expressions for the form factors for the different spin-singlet and -triplet symmetries possible on NN bonds (lowest harmonics). The overall amplitude is set to $\Delta_0$ and the distance between two NN carbon atoms is set to 1.}
\label{table1}
\end{table}

For a multilayer graphene with SC the pairing is intralayer and the Hamiltonian is given by 
\begin{eqnarray}
H=\sum_{\ell=1}^L \left(H_0^{\ell}+H_\text{NN}^{\eta,\ell}\right)+H_\text{interlayer}~,
\end{eqnarray}
where $L$ is the number of layers, $H_0^{\ell}$ and $H_\text{NN}^{\eta,\ell}$ are the non-interacting Hamiltonian and the SC Hamiltonian associated with each layer $\ell$, respectively, and given by expressions equivalent to Eqs.~(\ref{H_0})--(\ref{H_NN}), while $H_\text{interlayer}$ is the coupling Hamiltonian between adjacent layers, see e.g.~Ref.~\onlinecite{Pangburn2022}. Here we retain three important parameters in $H_\text{interlayer}$: (1) the interlayer hopping $\gamma_1$ between atoms directly on top of each other, (2) the interlayer hopping $\gamma_3$ between an atom $A$ in one layer and the neighboring $B$ atoms in the adjacent layer \cite{Malard2007} also denoted as trigonal warping, often neglected, although of the same order of magnitude as $\gamma_1$, and (3) a putative phase difference $\phi$ between the SC order parameters in two adjacent layers.

For the calculations, the reciprocal space Hamiltonian $H_{\bf k}$ in each layer $l$ is written in the basis 
\begin{equation}
 \mathcal{B}_l=\left(a_{kl\uparrow},b_{kl\uparrow},a_{kl\downarrow},b_{kl\downarrow},a^\dagger_{-kl\uparrow},b^\dagger_{-kl\uparrow},a^\dagger_{-kl\downarrow},b^\dagger_{-kl\downarrow}\right)^T~,
\end{equation}
such that
\begin{equation}\label{HBdG}
 H_{\bf k} = \frac{1}{2} \mathcal{B}^\dagger H_{\rm BdG} \mathcal{B}~,
\end{equation}
where $\mathcal{B}$ is the basis combining all individual-layer bases $\mathcal{B}_l$, and $H_{\rm BdG}$ is the $8L \times 8L$ Bogoliubov-de-Gennes (BdG) Hamiltonian matrix. The factor 8 corresponds to a product of $2$ spins, $2$ sublattices, and the particle-hole doubling of the degrees of freedom in the BdG formalism. Since there are no spin-flip terms in the graphene systems of interest, all bands are spin degenerate. Thus, in what follows all edge and impurity states are always doubly degenerate.

\subsection{Chern number}

To calculate the Chern number for the various SC states, we use the following expression\cite{Thouless1982,Ghosh2010,Sedlmayr2017}
\begin{eqnarray}\label{chern_def}
 \mathcal{C}&=&\frac{i}{8\pi^2}\int \text{d}{\bf k}\,\text{d}E\; \text{Tr}\bigg[G^2({\bf k},E)(\partial_{k_y}H_{\bf k})G({\bf k},E)(\partial_{k_x}H_{\bf k})\nonumber\\
 &&-G^2({\bf k},E)(\partial_{k_x}H_{\bf k})G({\bf k},E)(\partial_{k_y}H_{\bf k})\bigg]~,
\end{eqnarray}
where $G({\bf k},E)=(iE-H_{\rm BdG})^{-1}$ is the Matsubara Green's function. We note that as a consequence of the spin degeneracy of each band, the Chern numbers in what follows are always doubled compared to studies in which this doubling has been resolved~\cite{Volovikbook,ReadGreen2000,BlackSchaffer2012Edge,BlackSchaffer2014,Awoga2017Domain}.

\subsection{Edge spectral function}

\begin{figure}[t]
\begin{center}
\includegraphics[height=2.8cm]{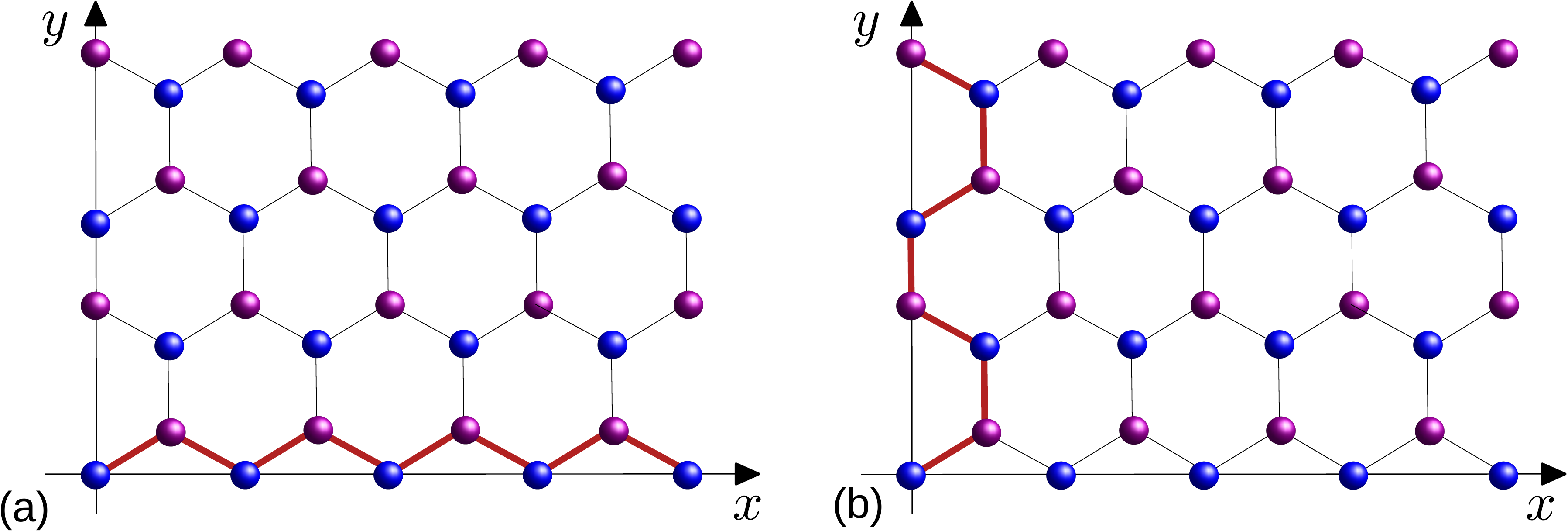}
\caption{Graphene lattice with A and B sites in different colors and with an impurity line (bold line) along (a)~a zigzag edge and (b) an armchair edge.}\label{figure_border}
\end{center}
\end{figure}

\begin{figure}[t]
\begin{center}
\includegraphics[height=2.8cm]{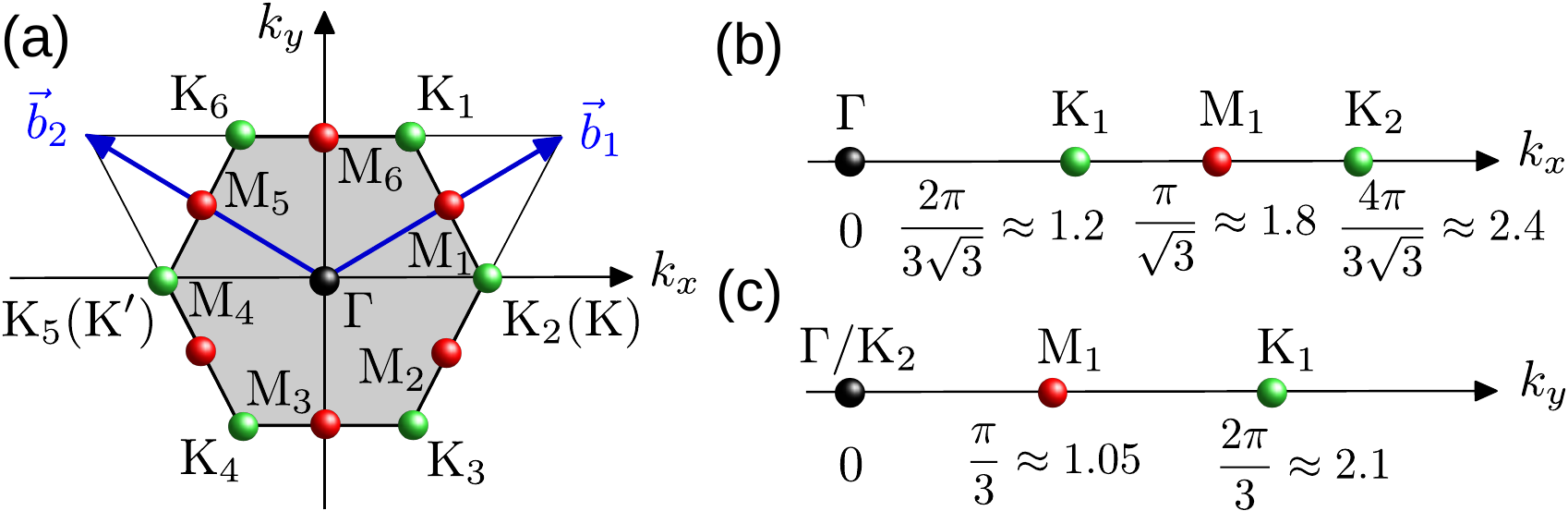}
\caption{(a) Schematic view of the first 2D Brillouin zone of the graphene honeycomb lattice with the 1D projections of the $\Gamma$ (black), $K$ (green), and $M$ (red) points on the $k_x$-axis in (b) and on the $k_y$-axis in (c). The $K_2$ and $K_5$ points correspond to what is traditionally denoted as the $K$ and $K^\prime$ points. Also depicted are the reciprocal unit vectors $\vec b_{1,2}=(\pm 2\pi/\sqrt{3},2\pi/3)$.}\label{figure_BZ}
\end{center}
\end{figure}

In honeycomb graphene lattices an edge can have different orientations. We restrict our study to the two most common, armchair and zigzag edges, depicted in Fig.~\ref{figure_border}. The energy spectrum of these edges can be extracted by considering a line-like impurity parallel to the edge which cuts the system into two semi-infinite systems. Scattering of electrons due to the presence of the impurity line induces additional features in the spatially-averaged electronic bulk spectral function $A({\bf k},E)$. The edge features are in this way captured as a correction $\delta A({\bf k},E)$ to the bulk spectral function, given by\cite{Pinon2020,Pinon2021},
\begin{eqnarray}
	&&\delta A(k_x,E)=-\frac{1}{\pi}\int_{-2\pi/3a}^{2\pi/3a}\frac{dk_y}{L_{BZ,y}}\nonumber\\
	&&\times\text{Im}\{\text{Tr}_\text{el}[G^r({\bf k},E)T(k_x,k_x,E)G^r({\bf k},E)]\}~,
\end{eqnarray}
 for the zigzag edge, with $L_{BZ,y}=4\pi/3a$, and
\begin{eqnarray}
 &&\delta A(k_y,E)=-\frac{1}{\pi}\int_{-2\pi/\sqrt{3}a}^{2\pi/\sqrt{3}a}\frac{dk_x}{L_{BZ,x}}\nonumber\\
 &&\times\text{Im}\{\text{Tr}_\text{el}[G^r({\bf k},E)T(k_y,k_y,E)G^r({\bf k},E)]\}~,
\end{eqnarray}
 for the armchair edge, with $L_{BZ,x}=4\pi/\sqrt{3}a$. 
Here the retarded Green's function is defined as $G^r({\bf k},E)=(E-H_{\rm BdG}+i0^+)^{-1}$.
The matrix trace $\text{Tr}_\text{el}$ indicates that the sum is performed only over the electron degrees of freedom. 
The T-matrix associated with a zigzag edge is defined as
\begin{eqnarray}\label{def_T1}
 T(k_x,k_x,E)=\left[\mathbb{1}_n-\mathbb{V}_n\int_{-2\pi/3a}^{2\pi/3a}\frac{dk_y}{L_{BZ,y}}G^r({\bf k},E)\right]^{-1}\mathbb{V}_n~,\nonumber\\
\end{eqnarray}
whereas for an armchair edge we have
\begin{eqnarray}\label{def_T2}
 T(k_y,k_y,E)=\left[\mathbb{1}_n-\mathbb{V}_n\int_{-2\pi/\sqrt{3}a}^{2\pi/\sqrt{3}a}\frac{dk_x}{L_{BZ,x}}G^r({\bf k},E)\right]^{-1}\mathbb{V}_n~.\nonumber\\
\end{eqnarray}
In Eqs.~(\ref{def_T1}) and (\ref{def_T2}), the identity matrix $\mathbb{1}_n$ and the impurity matrix $\mathbb{V}_n$ are $n\times n$ matrices with $n=8L$, with the latter given in the same basis as Eq.~(\ref{HBdG}) as 
\begin{eqnarray}
	\mathbb{V}_n=U\left(
	\begin{array}{cc}
		1&0\\
		0 & -1\\
	\end{array}
	\right) \otimes \mathbb{1}_{n/2}~,
\end{eqnarray}
where $U$ is the value of the potential on the impurity line. We set $U=1000t$ throughout this work to numerically simulate a hard edge. 
The calculations of edge spectrum occurs in the first two-dimensional (2D) Brillouin zone which is illustrated in Fig.~\ref{figure_BZ}, including 1D projections for each edge direction.

%%%%%%%%%%%%%%%%%%%%%%%%%%%%%%%%%%%%%%%%%%%%%%%%%%%%%%%%%%%%%%%%%%%%%%%%%%%%%%%%%%%%%%%%%%%%%%%%%%%%%%%%%%%%
%                                                                                                          %
%                                                                                                          %
%                           RESULTS MONOLAYER                                       %
%                                                                                                          %
%                                                                                                          %
%%%%%%%%%%%%%%%%%%%%%%%%%%%%%%%%%%%%%%%%%%%%%%%%%%%%%%%%%%%%%%%%%%%%%%%%%%%%%%%%%%%%%%%%%%%%%%%%%%%%%%%%%%%%

\section{Monolayer graphene}

\subsection{Chern number}
\begin{figure}[h!]
\begin{tikzpicture}
\node at (0,0) {
\includegraphics[width=4.2cm]{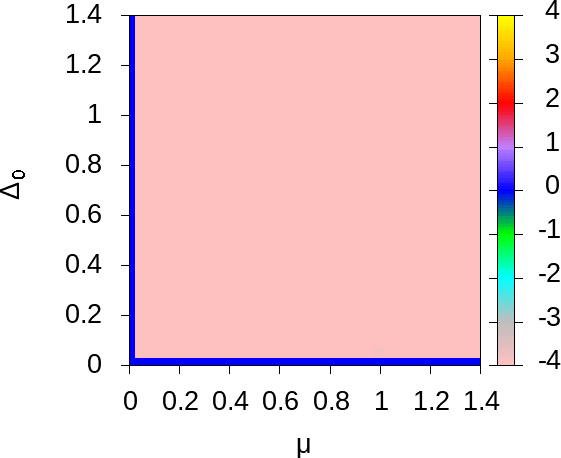}
\includegraphics[width=4.2cm]{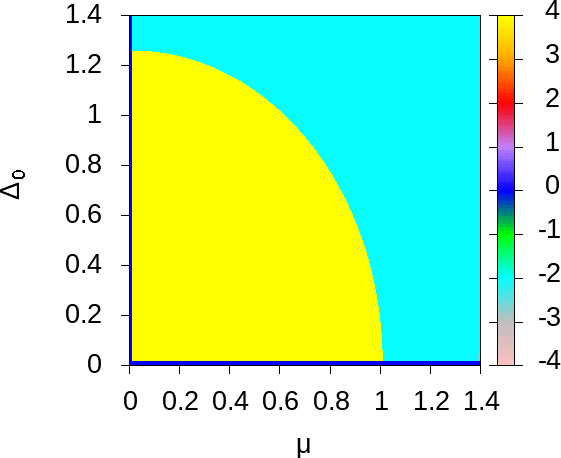}
};
\node at (-4.1,1.5) {(a)};
\node at (0.3,1.5) {(b)};
\end{tikzpicture}
\caption{Chern number $\mathcal{C}$ as a function of $\mu$ and $\Delta_0$ for monolayer graphene for the (a) $d+id'$- and (b) $p+ip'$-wave state.}
\label{figure_chern_monolayer}
\end{figure}

We start by studying monolayer graphene and first calculate the Chern number $\mathcal{C}$ using Eq.~(\ref{chern_def}). The calculation shows that~$\mathcal{C}$  is non-zero for both the $d+id'$- and  $p+ip'$-wave states, whereas it is equal to zero for all the other SC states, listed in Table \ref{table1}, including the $d_{xy}$-wave state. For the $d+id'$-wave state, the Chern number takes a constant value equal to $-4$ as seen in Fig.~\ref{figure_chern_monolayer}(a), for all values of chemical potential $\mu$ and superconducting order parameter amplitude $\Delta_0$, which are the two tunable parameters. More interestingly, for the $p+ip'$-wave state we observe a phase transition at large values of either $\mu$ or $\Delta_0$, as seen in Fig.~\ref{figure_chern_monolayer}(b). As shown in Ref.~\onlinecite{Pangburn2022}, this transition between a Chern number of $4$ (yellow) and $-2$ (cyan) corresponds to a gap closing line. The equation of this line is given by $\Delta_0 \approx \Delta_c\sqrt{1-(\mu/\mu_c)^2}$, with the critical parameters $\Delta_c\approx1.25$ and $\mu_c\approx1$ (in units of $t$). 

The behavior of the Chern number in Fig.~\ref{figure_chern_monolayer} can be understood starting from an analysis of the normal state band structure combined with the winding number of the SC order parameter~\cite{Volovikbook,BlackSchaffer2014}. At $\mu<t$ and small $\Delta_0$, there exist two Fermi surfaces centered around the $K$- and $K^\prime$-points, the Dirac points (see Appendix \ref{appa}). The $p+ip'$-wave order parameter winds once around each of these Fermi surfaces \cite{Pangburn2022}, thus giving a Chern number of 2 for each spin species (1 per Fermi surface), hence a total Chern number of 4. However, at $\mu>t$ and small $\Delta_0$, the normal state Fermi surface undergoes a Lifshitz transition and there is then only one Fermi surface centered at the $\Gamma$-point. The $p+ip'$-wave order parameter winds once per spin species around this central Fermi surface, giving a total Chern number of $-2$ (the sign of the Chern also changes at the Lifshitz transition). This analysis holds for small $\Delta_0$. When $\Delta_0$ become substantial, the border between the different Chern number regions changes, following the gap closing region established in Ref. \onlinecite{Pangburn2022}.

For the $d+id'$-wave state, the SC order parameter winds twice around the  $\Gamma$-centered Fermi surface due to its angular momentum, leading to a Chern number of $-4$ for $\mu>t$. Here the sign of the Chern number is the same as for the $p+ip'$-wave state at $\mu>t$ since they have the same winding direction. However, for $\mu<t$, the $d+id'$-wave state looks locally around the $K$- and $K^\prime$-points the same as the $p+ip'$-wave one \cite{BlackSchaffer2014,Pangburn2022}, except for an opposite winding at the two Dirac points. As a consequence, each Fermi surface and spin species contributes a (negative) unit to the Chern number, resulting in a $\mathcal{C} = -4$ for the chiral $d+id'$-wave state also at $\mu<t$, and no corresponding phase transition between the large $\mu$ and the small $\mu$ region. 

Based on this analysis it is clear that not only the winding of the SC order parameter determines the Chern number in monolayer graphene, but also that the normal state band structure and its Fermi surface are crucial for understanding the results in Fig.~\ref{figure_chern_monolayer}.

Having established the Chern numbers for all NN pairing states in monolayer graphene, we turn to examining the edge state spectrum, as visualized by the correction to the spectral function due to a line-like impurity. We first report data for the chiral $d+id'$-wave and $p+ip'$-wave states, and then finally also for the $d_{xy}$-wave state. We compare the number of observed edge states with the Chern number $\mathcal{C}$.

\subsection{Edge states for $d+id'$-wave symmetry}

\begin{figure}[t]
{\sc\footnotesize \hspace*{0.3cm}Armchair edges\hspace{2cm}Zigzag edges}
\begin{tikzpicture}
\node at (0,0) {
\includegraphics[width=4.2cm]{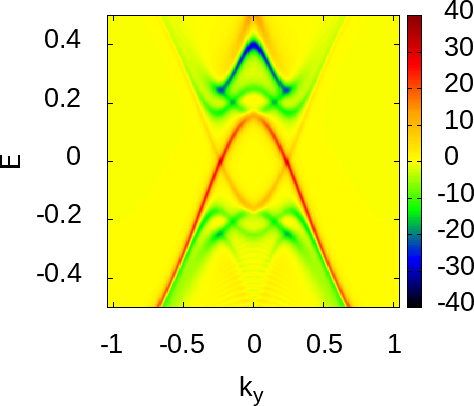}
\includegraphics[width=4.2cm]{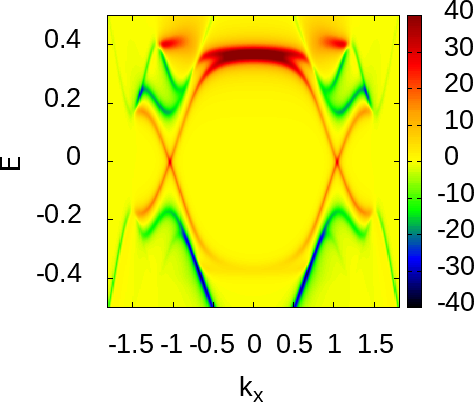}
};
\node at (-4.1,1.5) {(a)};
\end{tikzpicture}
\begin{tikzpicture}
\node at (0,0) {
\includegraphics[width=4.2cm]{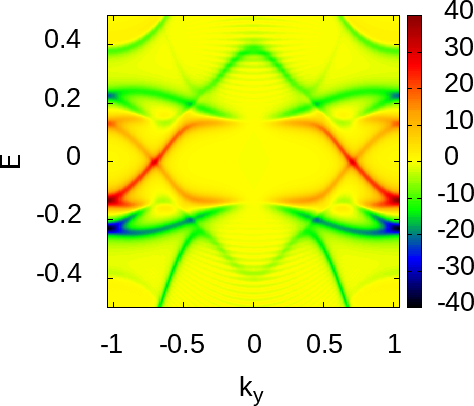}
\includegraphics[width=4.2cm]{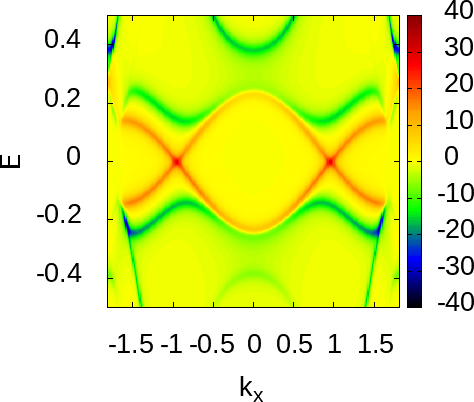}
};
\node at (-4.1,1.5) {(b)};
\end{tikzpicture}
\caption{Edge correction to the spectral function for monolayer graphene with a spin-singlet $d+id'$-wave SC order parameter with $\Delta_0=0.4$ at (a)~$\mu=0.4$  and (b)~$\mu=1.2$. The corresponding Chern number is $\mathcal{C}=-4$.}\label{figure_d+id'_monolayer}
\end{figure}

For the chiral $d+id'$-wave SC state, Fig.~\ref{figure_chern_monolayer}(a) shows that the Chern number $\mathcal{C}$ takes a constant value equal to $-4$ for any $\mu$ and non-zero values of $\Delta_0$. This means that four edge states are expected at each edge, which with the spin degeneracy makes 2 distinguishable states.\cite{BlackSchaffer2012Edge,Awoga2017Domain} This is verified in Fig.~\ref{figure_d+id'_monolayer} where we plot the correction to the spectral function due to a line-like impurity modeling the edge for both armchair and zigzag edges for two different values of $\mu$ at $\Delta_0=0.4$. We observe two main characteristics: a hard energy gap of width of the order of $\Delta_0$ opens in the spectrum and, inside this gap, a total of 4 states appear (red color). These states are all doubly degenerate resulting in a total of 8 edge states: we have checked that these states are spin degenerate by adding a Zeeman field, indeed the number of edge states doubled in the presence of magnetic field, confirming their spin degeneracy.  With the Chern number $\mathcal{C}=-4$, the appearance of 8 edge states may at first seems to be a discrepancy, but we then note that the correction to the spectral function plotted here actually corresponds to a spatial average over the entire system, thus taking into account two edges, one on each side of the impurity line. We thus always plot the edge spectrum for two edges. In this case this gives rise to two spin-degenerate chiral and co-propagating edge states on each edge, while the edge states on different edges are counter-propagating, as expected from the chirality of the SC order parameter. As a consequence, we find a full agreement between the bulk Chern number and the boundary edge states, fulfilling the bulk-boundary correspondence.

%We also note that the zero-crossing of the edge states can occur at different $k$-values for $\mu<t$ (a) and $\mu>t$ (b) as well as different edge terminations. This is because the edge states will appear in the region with the lowest bulk energy gap. The bulk energy gap is set by both the normal state band structure and the SC order parameter, and is generally small around the normal-state Fermi surface. This means the minimum energy gap is found centered around the $K$ and $K^\prime$-points for $\mu<t$ and around the $\Gamma$-point for $\mu>t$. Different projections of the 2D $K$ and $K^\prime$-momenta into the edge state 1D momenta for the two different edges, see Fig.~\ref{figure_BZ}, give rise to the different zero-energy crossing momenta for the two different edges in (a). However, the $\Gamma$-point remains at zero momentum for both edges, and the edge state spectrum thus appear at similar $k$-values in (b) and as a consequence look remarkably similar.

\subsection{Edge states for $p+ip'$-wave symmetry}
\begin{figure}[t]
{\sc\footnotesize \hspace*{0.3cm}Armchair edges\hspace{2cm}Zigzag edges}
\begin{tikzpicture}
\node at (0,0) {
\includegraphics[width=4.2cm]{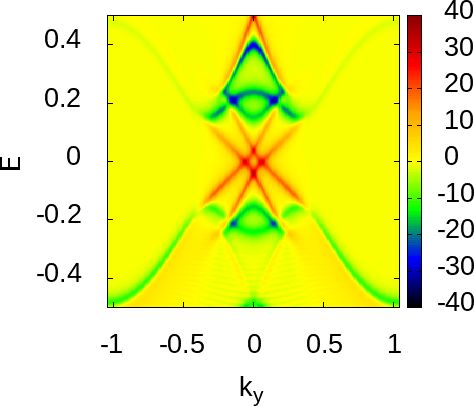}
\includegraphics[width=4.2cm]{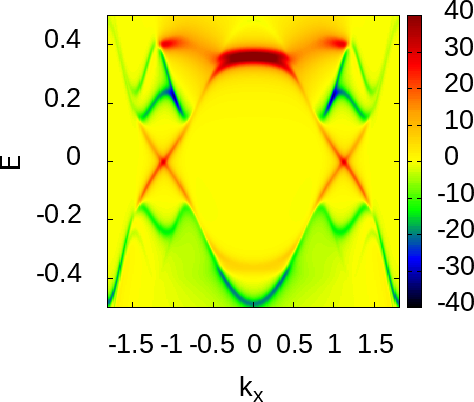}
};
\node at (-4.1,1.5) {(a)};
\end{tikzpicture}
\begin{tikzpicture}
\node at (0,0) {
\includegraphics[width=4.2cm]{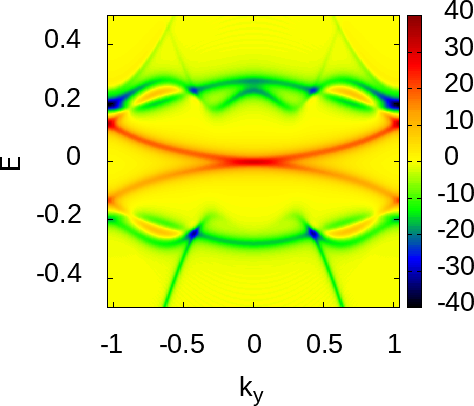}
\includegraphics[width=4.2cm]{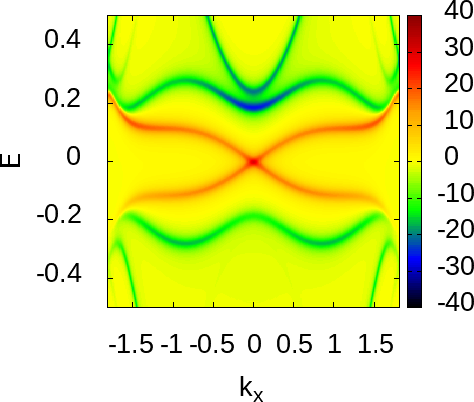}
};
\node at (-4.1,1.5) {(b)};
\end{tikzpicture}
\caption{Edge correction to the spectral function for monolayer graphene with a $p+ip'$-wave SC order parameter with $\Delta_0=0.4$ at (a)~$\mu=0.4$ corresponding to $\mathcal{C}=4$, and (b)~$\mu=1.2$  corresponding to $\mathcal{C}=-2$.}
\label{figure_p+ip'_monolayer}
\end{figure}

For a $p+ip'$-wave state the correction to the spectral function due to an impurity line along either the armchair or zigzag edge is shown in Fig.~\ref{figure_p+ip'_monolayer}. Similarly to the $d+id'$-wave state, we observe once more the opening of a full energy gap of width of the order of $\Delta_0$ and, inside this gap, the formation of edge states. For $\Delta_0=\mu=0.4$, we count four spin-degenerate edge states in total for the two present edges, see Fig.~\ref{figure_p+ip'_monolayer}(a). On each edge we have 2 spin-degenerate co-propagating states, with states counter-propagating between edges.  Increasing to $\mu=1.2$ we instead find two such spin-degenerate edge states, see Fig.~\ref{figure_p+ip'_monolayer}(b). Both of these results are in full agreement with the values taken by $\mathcal{C}$, which is equal to $4$ in (a) and $-2$ in (b). 

%By looking more in detail  at Fig.~\ref{figure_p+ip'_monolayer}, we see that for $\mu>t$ (b) the edge states for zigzag and armchair edges both cross at $k_{x,y}=0$, which corresponds to the position of the projections of the $\Gamma$-point along $k_x$ and $k_y$, respectively. 
We also note that the zero-crossing of the edge states can occur at different $k$-values for $\mu<t$ (a) and $\mu>t$ (b) as well as different edge terminations. This is because the edge states will appear in the region with the lowest bulk energy gap. The bulk energy gap is set by both the normal state band structure and the SC order parameter, and is generally small around the normal-state Fermi surface. This means the minimum energy gap is found centered around the $K$ and $K^\prime$-points for $\mu<t$ and around the $\Gamma$-point for $\mu>t$. Different projections of the 2D $K$ and $K^\prime$-momenta into the edge state 1D momenta for the two different edges: $k_y=0$, the projection of $K/K^\prime$ for armchair in  Fig.~\ref{figure_BZ} c), and $k_x=2\pi/3 \sqrt{3}\approx 1.2$, the projection of $K_1$ in Fig.~\ref{figure_BZ} b) for zigzag, give rise to the different zero-energy crossing momenta for the two different edges in (a). However, the projection of the $\Gamma$-point remains at zero momentum for both edges, and the edge state spectra thus appear in the vicinity of similar $k=0$-values in (b), and as a consequence the edge states for armchair and zigzag look remarkably similar when $\mu>t$.
%For $\mu<t$ (a) the edge states instead cross at very different $k$-values due to the different projections fo the $K$ and $K^\prime$-points in to the 1D Brillouin zones for each edge, just the same as for $d+id'$-wave pairing in Fig.~\ref{figure_d+id'_monolayer}. 
Thus, by increasing the value of the chemical potential, the crossing point of the edge states moves between the $K$ and $K^\prime$-points to the $\Gamma$-point. 

\subsection{$d_{xy}$-wave symmetry}\label{subsec:dxy}
\begin{figure}[t]
{\sc\footnotesize \hspace*{0.3cm}Armchair edges\hspace{2cm}Zigzag edges}
\begin{tikzpicture}
\node at (0,0) {
\includegraphics[width=4.2cm]{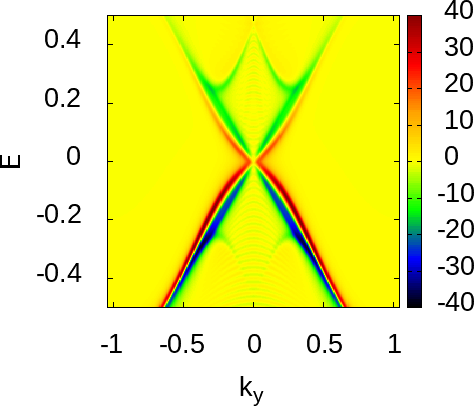}
\includegraphics[width=4.2cm]{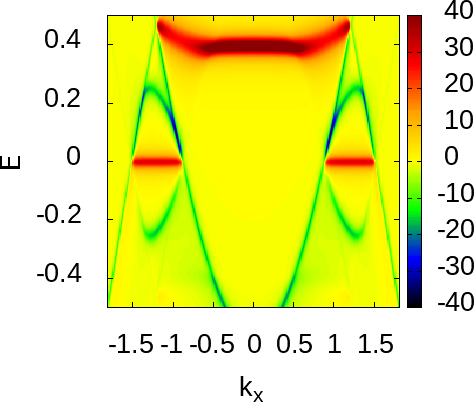}
};
\node at (-4.1,1.5) {(a)};
\end{tikzpicture}
\begin{tikzpicture}
\node at (0,0) {
\includegraphics[width=4.2cm]{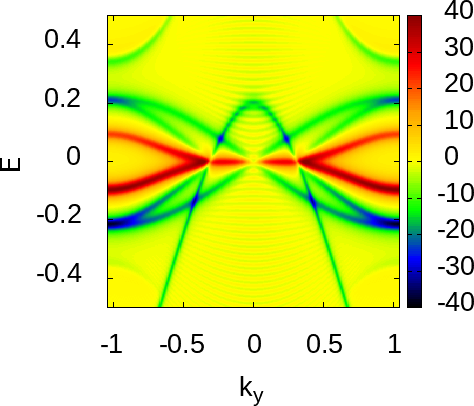}
\includegraphics[width=4.2cm]{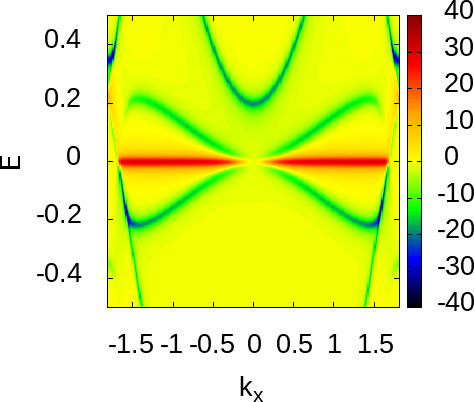}
};
\node at (-4.1,1.5) {(b)};
\end{tikzpicture}
\caption{Edge correction to the spectral function of monolayer graphene with a $d_{xy}$-wave SC order parameter with $\Delta_0=0.4$ at (a)~$\mu=0.4$ and (b)~$\mu=1.2$. This corresponds to a bulk $\mathcal{C}=0$}\label{figure_dxy_monolayer}
\end{figure}

Finally we also report the edge spectrum for the $d_{xy}$-wave SC state where the Chern number $\mathcal{C} =0$, independent of the values of $\mu$ and $\Delta_0$, and the system is also gapless \cite{Pangburn2022}. Figure~\ref{figure_dxy_monolayer} displays the correction to the spectral function in the presence of an armchair edge or zigzag edge, again for $\mu<t$ and $\mu>t$. Here we first note that the bulk is gapless as there exists green/blue lines (i.e.~suppressed edge spectrum) all the way down to zero energy. Second, we see the formation of edge states (red), which in several cases consist of zero-energy flat bands. These are not of the same chiral topological origin as the edge states in the chiral $d+id'$- and $p+ip'$-wave superconductors in Figs.~\ref{figure_d+id'_monolayer}-\ref{figure_p+ip'_monolayer}, but instead edge states found on certain surfaces in nodal superconductors. The most known example is the zero-energy flat band existing on the [110] edge of the cuprate $d$-wave superconductors \cite{Kashiwaya00,Lofwander01}. 
These are Andreev bound states at zero-energy and existing for all $k$-values in the edge 1D Brillouin zone bounded by the projections of the bulk superconducting nodes.\cite{Ryu02,Sato11, Potter14, Chakraborty22} For the cuprate $d$-wave superconductors this means a flat band of zero-energy edge states for the [110] edge, but not at the [100] or [010] edges. While these edge states are also protected by a topological argument,\cite{Potter14} they are much more fragile than those protected by a finite Chern number, and can e.g.~be destroyed by edge inhomogeneity. For the zigzag edge we see similarly to the cuprates that a flat band of zero-energy edge states (red) exist between the two bulk nodal points (green). However, the armchair edge at $\mu<t$ lacks a zero-energy flat band since there is no distance between the 2 bulk nodal points when they are both projected onto the 1D armchair edge Brillouin zone as the bulk nodal points both occur at $k_y=0$, see Fig.~\ref{figure_BZ}. However, the armchair edge for $\mu>t$ has a finite projection, due to a total of 4 bulk nodal points for a Fermi surface around $\Gamma$, and we also now find a zero-energy edge flat band.

%%%%%%%%%%%%%%%%%%%%%%%%%%%%%%%%%%%%%%%%%%%%%%%%%%%%%%%%%%%%%%%%%%%%%%%%%%%%%%%%%%%%%%%%%%%%%%%%%%%%%%%%%%%%
%                                                                                                          %
%                                                                                                          %
%                           RESULTS BILAYER                                               %
%                                                                                                          %
%                                                                                                          %
%%%%%%%%%%%%%%%%%%%%%%%%%%%%%%%%%%%%%%%%%%%%%%%%%%%%%%%%%%%%%%%%%%%%%%%%%%%%%%%%%%%%%%%%%%%%%%%%%%%%%%%%%%%%

\section{Bilayer graphene}

Given that the $p+ip'$-wave state is the only state which exhibits a phase transition between different Chern number regions for monolayer graphene, we choose in the following to focus primarily on this state for the bilayer and trilayer graphene systems. However, in Appendix~\ref{appb}, we provide some results obtained for bilayer and trilayer graphene with a $d+id'$-wave SC order parameter.

First, we consider an AB-stacked bilayer graphene system characterized by a $p+ip'$-wave SC order parameter, with an interlayer hopping $\gamma_1$, both in the absence and presence of trigonal warping $\gamma_3$. We first calculate the Chern number as a function of the chemical potential $\mu$ and the SC order parameter amplitude $\Delta_0$, and we subsequently also study the formation of edge states.

\begin{figure}[t]
\begin{tikzpicture}
\node at (0,0) {
\includegraphics[height=3.5cm]{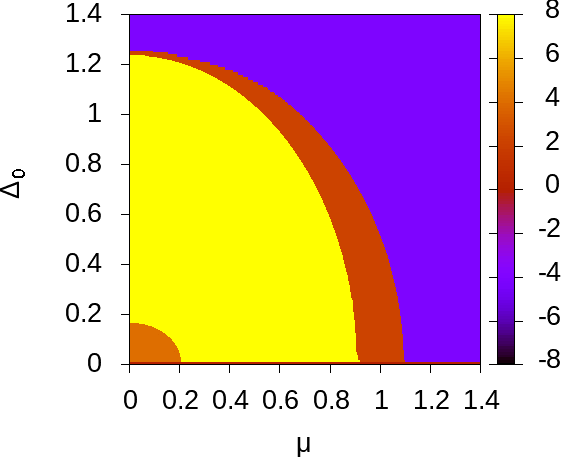}
\includegraphics[height=3.5cm]{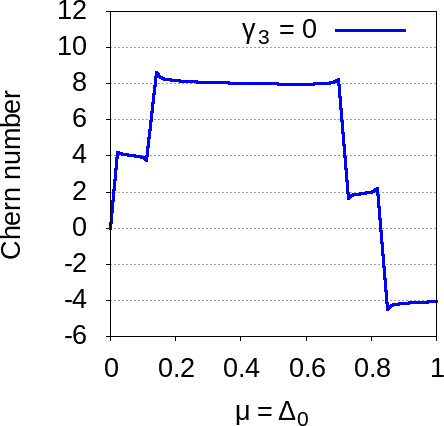}
};
\node at (-4,1.5) {(a)};
\end{tikzpicture}
\begin{tikzpicture}
\node at (0,0) {
\includegraphics[height=3.5cm]{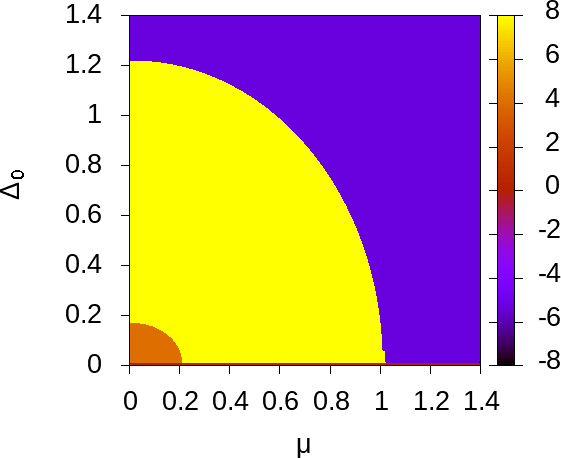}
\includegraphics[height=3.5cm]{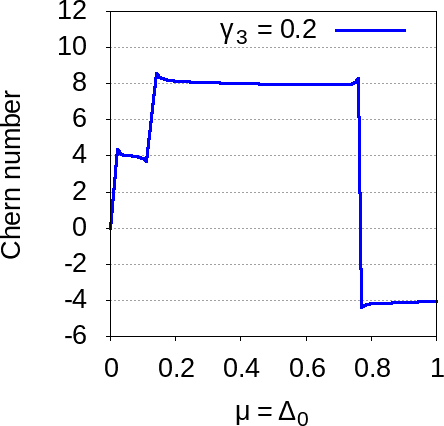}
};
\node at (-4,1.5) {(b)};
\end{tikzpicture}
\caption{(Left) Chern number $\mathcal{C}$ for AB-bilayer graphene with a $p+ip'$-wave SC order parameter as a function of $\mu$ and $\Delta_0$ and (Right) Chern number along the diagonal $\mu=\Delta_0$ for $\gamma_1=0.2$ and (a)~without trigonal warping $\gamma_3=0$ and (b)~with trigonal warping $\gamma_3=0.2$.}\label{figure_chern_bilayer}
\end{figure}

\subsection{Bilayer graphene Chern number}
In Fig.~\ref{figure_chern_bilayer} we plot the Chern number $\mathcal{C}$ for bilayer graphene with $p+ip'$-wave SC order parameter, both with and without trigonal warping, $\gamma_3$. For $\gamma_3=0$ we find that phase diagram for the Chern number as a function of $\mu$ and $\Delta_0$ contains four different regions as depicted in Fig.~\ref{figure_chern_bilayer}(a): an orange region with $\mathcal{C}=4$, a yellow region with $\mathcal{C}=8$, a red region with $\mathcal{C}=2$ and finally a purple region with $\mathcal{C}=-4$. The right plot shows additionally a line-cut at $\mu =\Delta_0$.
In the presence of warping, i.e.~for $\gamma_3\ne 0$, the red region disappears and we end up with only three regions in the phase diagram, see Fig.~\ref{figure_chern_bilayer}(b). The boundaries between the regions with different values of~$\mathcal{C}$ can be described by the functional form $\Delta_0\approx\Delta_c\sqrt{1-(\mu/\mu_c)^2}$ with critical parameters $\Delta_c$ and $\mu_c$ extracted by fitting and given in Table~\ref{table2}. These boundaries correspond exactly to the gap closing lines described in Ref.~\onlinecite{Pangburn2022}, and their dependence on the trigonal warping is described in more detail in this previous work.

\begin{table}[h!]
\begin{tabular}{|l|c|c|}
  \hline
 &$\quad\Delta_c\quad$ & $\quad\mu_c\quad$ \\
   \hline
   Monolayer ($\gamma_1=\gamma_3=0$)  & 1.25 & 1\\
   \hline
Normal bilayer  ($\gamma_1=0.2$ and $\gamma_3=0$) & 
 \begin{tabular}{c}
 0.16\\1.23\\1.29
\end{tabular}
 & 
  \begin{tabular}{c}
 0.2\\0.9\\1.1
\end{tabular}\\
 \hline
Warped bilayer  ($\gamma_1=\gamma_3=0.2$) & 
 \begin{tabular}{c}
 0.16\\1.21
\end{tabular}
&
 \begin{tabular}{c}
 0.2\\1
\end{tabular}
\\
  \hline
\end{tabular}\caption{Values of the critical parameters $\Delta_c$ and $\mu_c$ for mono- and bi-layer graphene for $p+ip'$ pairing.}
\label{table2}
\end{table}

We can understand the results in Fig.~\ref{figure_chern_bilayer} by again analyzing the normal state band structure, see Appendix \ref{appa} for details, as for the monolayer case. First we ignore $\gamma_3$ and focus on small $\Delta_0$. Then, at the $K$ and $K^\prime$-points, the two bands in bilayer graphene are separated by an energy $E~\sim \gamma_1$. Thus, for $\mu<\gamma_1$, which is the region most likely to be accessed experimentally, there is only one band contributing to the Chern number, yielding $\mathcal{C}=4$, same as for monolayer graphene ($2$ spin species and $2$ Fermi surfaces).  At $\gamma_1 \le \mu <t-\gamma_1/2 $, both bands are filled and thus contribute to the Chern number, which jumps to $8$. In the vicinity of the Lifshitz transition we have now also a splitting of the bands. As a consequence, for $t-\gamma_1/2\le \mu \le t+\gamma_1/2$  we have one band which contributes with two Fermi surfaces centered around the $K$ and $K^\prime$ points and thus a total of $\mathcal{C}=4$, while a second band has its Fermi surface centered around $\Gamma$ and contributes a $\mathcal{C}=-2$. The total Chern is thus expected to be $\mathcal{C}=4-2=2$, as also seen in Fig.~\ref{figure_chern_bilayer}(a). Next turning on the trigonal warping $\gamma_3\neq 0$, there is a splitting of the bands at the $M$ points and hence the trigonal warping strongly influences the size of the $\mathcal{C}=2$ region. In fact, if $\gamma_1\approx \gamma_3$ this region shrinks to zero. Finally, for $\mu>t+\gamma_1/2$ both bands have a Fermi surface centered around $\Gamma$, and we find $\mathcal{C}=-4$. For larger $\Delta_0$ the above transition points follows the gap closing lines established in Table \ref{table2}.
 
\begin{figure}[t]
 \includegraphics[height=3.3cm]{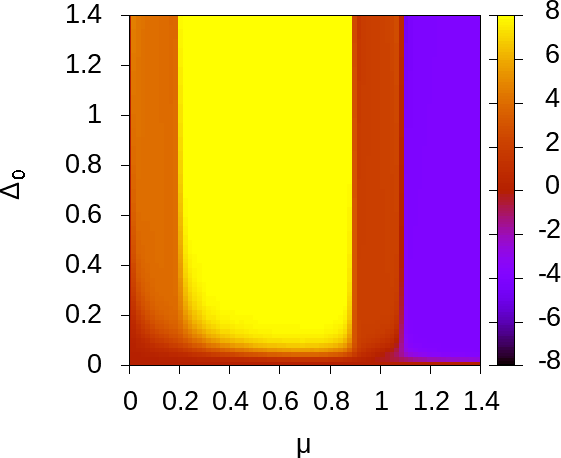}
\includegraphics[height=3.3cm]{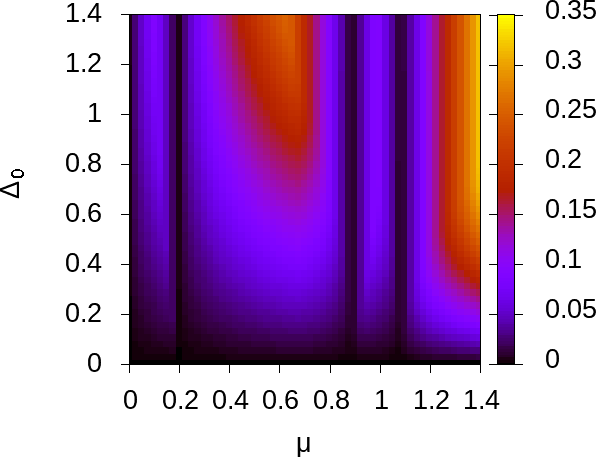}
 \caption{(a) Chern number and (b) energy gap for AB-bilayer graphene with a $p+ip'$-wave SC order parameter on NNN bonds as a function of $\mu$ and $\Delta_0$ for $\gamma_1=0.2$, $\gamma_3=0$. 
 \label{figure_chern_bilayer_dephased_NNN}}
\end{figure}

Before moving on with studying the edge spectrum, we make two interesting extensions. First, we would like to understand how these above results are modified if the $p+ip'$-wave state stems from a NNN pairing range rather from the used NN pairing. This is important to address since NNN range may be preferred over the NN coupling, especially in multi-layer graphene.\cite{chou2021acoustic,AwogaABC} The details for the Hamiltonian with an NNN SC order parameter are provided in Ref.~\onlinecite{Pangburn2022}. The gap closing analysis in Ref.~\onlinecite{Pangburn2022} indicates that the main difference between the NNN and NN pairing is a gap closing as a function of $\mu$, but notably independent of $\Delta_0$. We thus expect also a phase diagram for which the Chern number is independent of $\Delta_0$. This is verified in  Fig.~\ref{figure_chern_bilayer_dephased_NNN} where we observe the same topological phases for NNN  pairing as for NN pairing: an orange region with $\mathcal{C}=4$, a yellow region with $\mathcal{C}=8$, a red region with $\mathcal{C}=2$, and a purple region with $\mathcal{C}=-4$, but these are notably only affected by changing $\mu$, not  by the value of $\Delta_0$. 

\begin{figure}[t]
\begin{tikzpicture}
\node at (0,0) {
\includegraphics[height=3.3cm]{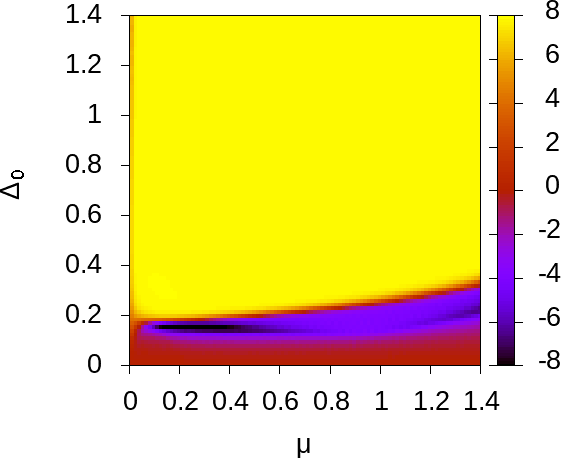}
\includegraphics[height=3.3cm]{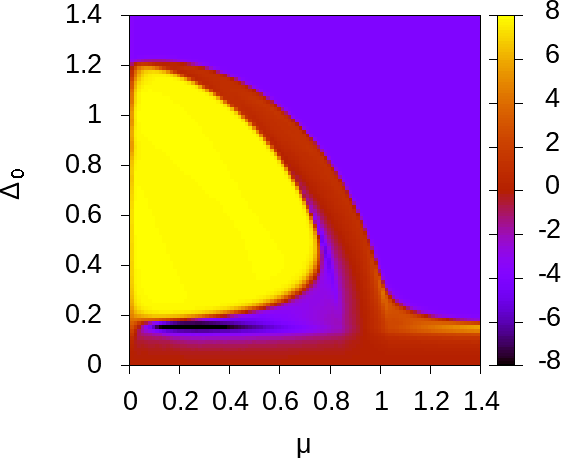}
};
\node at (-4.1,1.5) {(a)};
\node at (0.3,1.5) {(b)};
\end{tikzpicture}
\caption{Chern number $\mathcal{C}$ for AB-bilayer graphene with (a) $d+id'$- and (b) $p+ip'$-wave SC order parameter as a function of $\mu$ and $\Delta_0$ at $\gamma_1=0.2$ and $\gamma_3=0$, with a phase difference $\phi=\pi$ between the SC orders in the two graphene layers.}
\label{figure_chern_bilayer_dephased}
\end{figure}

Second, it is possible that a phase $\phi=\pi$ arises in certain situations between the SC order parameters in different graphene layers. Such phase difference has for example been noted in twisted bilayer graphene.\cite{Wu2019,Fischer2021,lothman2022nematic} While no such phase difference has been observed in regular graphene we believe it is still interesting here to understand its role for the topological phase diagram. Thus in Fig.~\ref{figure_chern_bilayer_dephased} we plot the value of the Chern number for  bilayer graphene with NN $d+id'$- and $p+ip'$-wave state when there is phase difference $\phi=\pi$ between the SC order parameters in the two graphene layers. We see that the main effect of this $\pi$-phase difference is the destruction of the topological phase at small values of $\Delta_0$, such that a trivial region with $\mathcal{C}=0$ is formed for small $\Delta_0$. This is particularly interesting to note, since it indicates that such a SC dephasing may destroy the topological properties of a multilayer system in the experimentally accessible regime.

\subsection{Bilayer graphene edge states}

\begin{figure}[t]
{\sc\footnotesize \hspace*{0.3cm}Armchair edges\hspace{2cm}Zigzag edges}
\begin{tikzpicture}
\node at (0,0) {
\includegraphics[width=4.2cm]{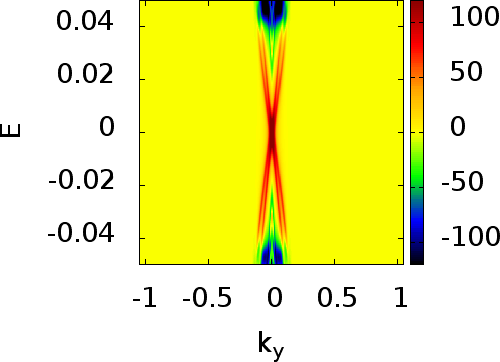}
\includegraphics[width=4.2cm]{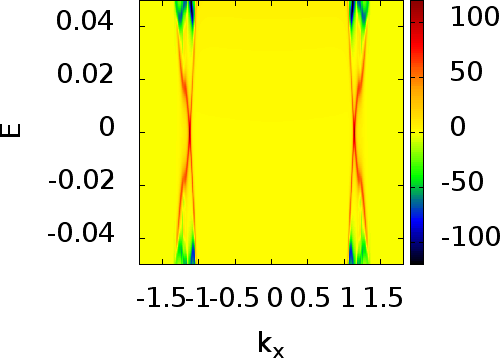}
};
\node at (-4.1,1.5) {(a)};
\end{tikzpicture}
\begin{tikzpicture}
\node at (0,0) {
\includegraphics[width=4.2cm]{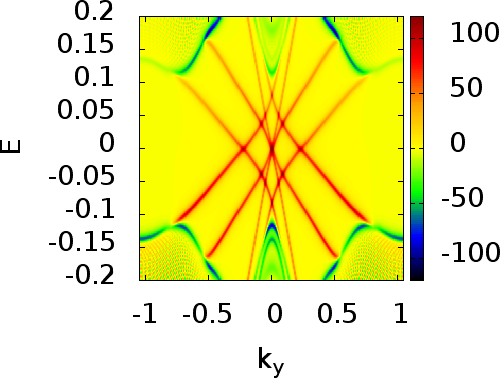}
\includegraphics[width=4.2cm]{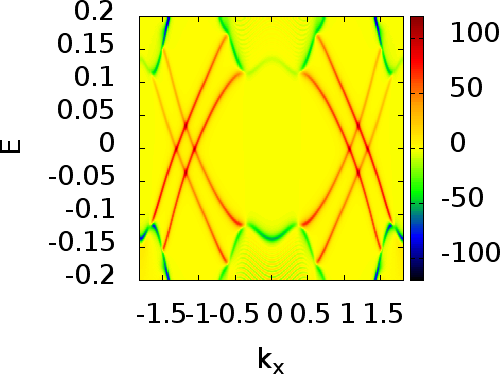}
};
\node at (-4.1,1.5) {(b)};
\end{tikzpicture}
\begin{tikzpicture}
\node at (0,0) {
\includegraphics[width=4.2cm]{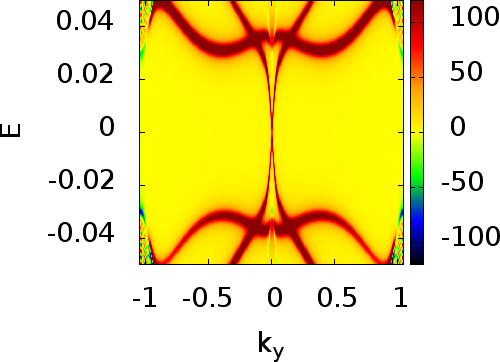}
\includegraphics[width=4.2cm]{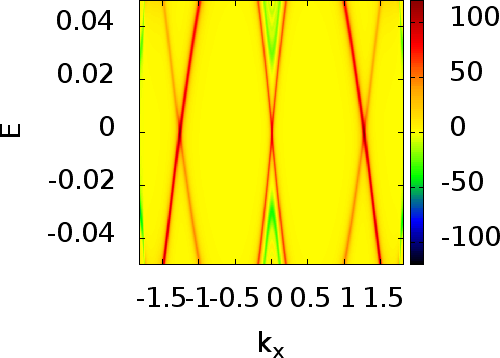}
};
\node at (-4.1,1.5) {(c)};
\end{tikzpicture}
\begin{tikzpicture}
\node at (0,0) {
\includegraphics[width=4.2cm]{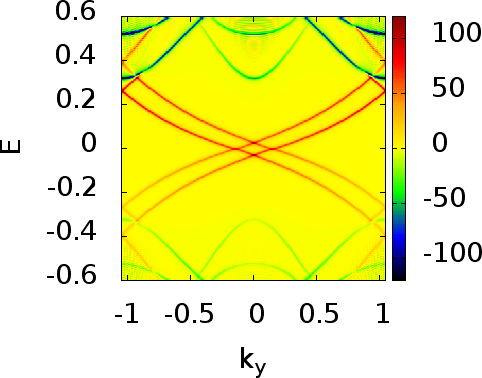}
\includegraphics[width=4.2cm]{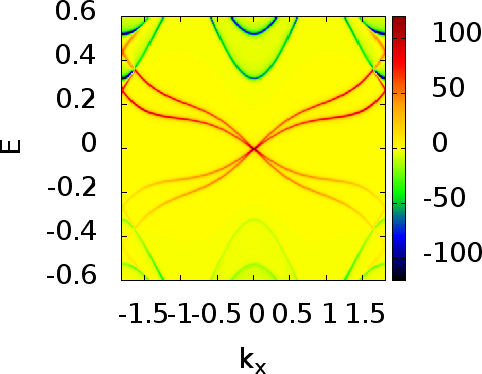}
};
\node at (-4.1,1.5) {(d)};
\end{tikzpicture}
\caption{Edge correction to the spectral function for AB-bilayer graphene with a $p+ip'$-wave SC order parameter with $\gamma_1=0.2$ and $\gamma_3=0$ at (a) $\Delta_0=\mu=0.1$  corresponding to $\mathcal{C}=4$, (b) $\Delta_0=\mu=0.6$  corresponding to $\mathcal{C}=8$, (c) $\Delta_0=\mu=0.8$ corresponding to $\mathcal{C}=2$, and (d) $\Delta_0=\mu=1.1$  corresponding to $\mathcal{C}=-4$.}
\label{figure_p+ip'_bilayer_without_warping}
\end{figure}

Moving on to the edge spectrum for bilayer graphene, we show in Fig.~\ref{figure_p+ip'_bilayer_without_warping}  the edge correction to the spectral function for a $p+ip'$-wave SC state without warping, $\gamma_3=0$, in the four different Chern number regions identified in Fig.~\ref{figure_chern_bilayer}(a). For the armchair edge, the number of edge states corresponds indeed to the absolute value of the Chern number since one counts 
four edge states in Fig.~\ref{figure_p+ip'_bilayer_without_warping}(a), 
eight edge states in Fig.~\ref{figure_p+ip'_bilayer_without_warping}(b), 
two edge states in Fig.~\ref{figure_p+ip'_bilayer_without_warping}(c), and 
four edge states in Fig.~\ref{figure_p+ip'_bilayer_without_warping}(d) which correspond to regions in the phase diagram inside which 
$\mathcal{C}$ is equal to $4$, $8$, $2$ and~$-4$,  respectively (all states are spin-degenerate but all plots also shows the edge spectrum for two edges with counter-propagating edge states, just as for monolayer graphene). 
However, for a zigzag edge one difference arises in Fig.~\ref{figure_p+ip'_bilayer_without_warping}(c), corresponding to values of $\mu$ and $\Delta_0$ that are located in the red $\mathcal{C} = 2$ region of Fig.~\ref{figure_chern_bilayer}(a). While $\mathcal{C}=2$, we can count six states crossing at zero energy. Such a situation has also been described in Ref.~\onlinecite{Sedlmayr2017}, where it was found that in some situations extra crossings may arise that are not topologically protected such that they can be removed by disorder or other modifications. Direct tight-binding calculations of the band structures show that this is what happens here as well and only the crossing at $k_x=0$ is topologically protected; each topologically protected band can in principle cross zero energy an odd number of times.
%{\bf We may eventually add a TB calculation here with and without disorder to check}.

%%%%%%%%%%%%%%%%%%%%%%%%%%%%%%%%%%%%%%%%%%%%%%%%%%%%%%%%%%%%%%%%%%%%%%%%%%%%%%%%%%%%%%%%%%%%%%%%%%%%%%%%%%%%
%                                                                                                          %
%                                                                                                          %
%                           RESULTS TRILAYER                                            %
%                                                                                                          %
%                                                                                                          %
%%%%%%%%%%%%%%%%%%%%%%%%%%%%%%%%%%%%%%%%%%%%%%%%%%%%%%%%%%%%%%%%%%%%%%%%%%%%%%%%%%%%%%%%%%%%%%%%%%%%%%%%%%%%

 % START HERE:
\begin{figure}[t]
%{\sc\footnotesize \hspace*{0.2cm}ABA-stacked\hspace{2.6cm}ABC-stacked}
\begin{tikzpicture}
\node at (0,0) {
\includegraphics[height=3.5cm]{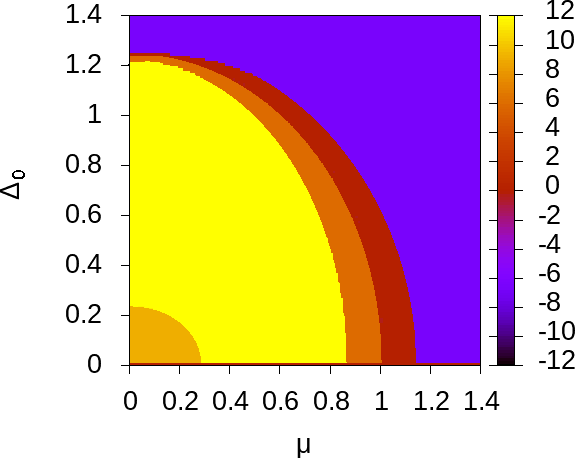}
\includegraphics[height=3.5cm]{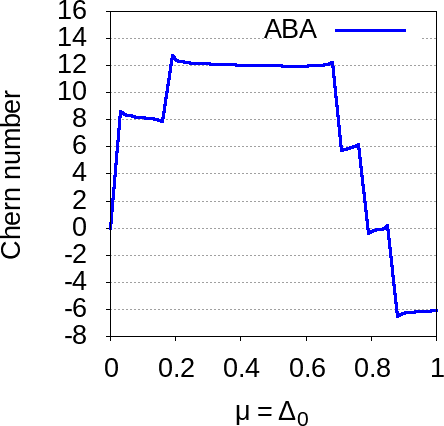}
};
\node at (-4,1.5) {(a)};
\end{tikzpicture}
\begin{tikzpicture}
\node at (0,0) {
\includegraphics[height=3.5cm]{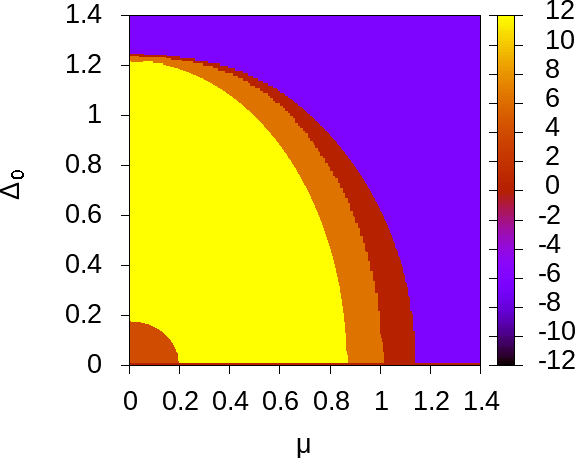}
\includegraphics[height=3.5cm]{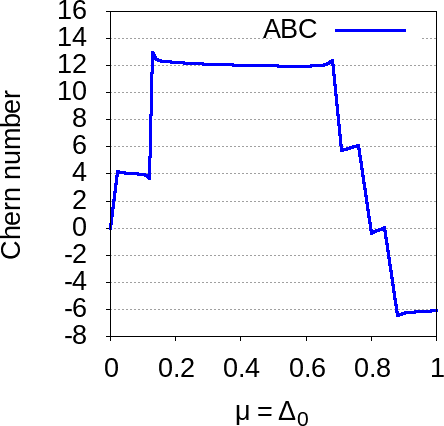}
};
\node at (-4,1.5) {(b)};
\end{tikzpicture}
\caption{(Left) Chern number $\mathcal{C}$ for (a) ABA-stacked and (b) ABC-stacked trilayer graphene with a $p+ip'$-wave SC order parameter as a function of $\mu$ and $\Delta_0$ and (Right) Chern number along the diagonal $\mu=\Delta_0$ for $\gamma_1=0.2$ and $\gamma_3=0$.}\label{figure_chern_trilayer}
\end{figure}

\section{Trilayer graphene}

Next we consider trilayer graphene with both  ABA- and ABC-stacking, again with a $p+ip'$-wave SC order parameter and with interlayer hopping $\gamma_1$. We calculate the Chern number and study the edge states first in the absence and then in the presence of trigonal warping $\gamma_3$.

\subsection{Trilayer graphene Chern number}

For ABA-stacked trilayer graphene without trigonal warping ($\gamma_3=0$), the phase diagram for the Chern number contains five different regions as depicted in Fig.~\ref{figure_chern_trilayer}(a): a light-orange region with $\mathcal{C}=8$, a yellow region with $\mathcal{C}=12$, a dark-orange region with $\mathcal{C}=6$, a red region with $\mathcal{C}=0$, and finally a purple region with $\mathcal{C}=-6$. For  ABC-stacked trilayer graphene without warping the phase diagram for the Chern number presented in Fig.~\ref{figure_chern_trilayer}(b) and has a similar appearance except that at low-energy we find $\mathcal{C}=4$ rather than $\mathcal{C}=8$. 
To understand the phase diagrams, we also perform similar calculations for tetralayer ($L=4$) and pentalayer ($L=5$) graphene systems, with results reported in Appendix \ref{appc}. We note that a Chern number of~$4$ for $\mu<\gamma_1$ is only possible for a fully rhombohedral stacking, i.e.~ABC for bilayer, ABCA for tetralayer and ABCAB for pentalayer graphene. The fact that the Chern number in this region is universally $\mathcal{C}=4$ and does not depend on the number of layers is quite a remarkable result, and could serve as a way to identify this type of stacking which has particularly different topological properties. 

Next we again note the influence of the normal state band structure. For small values of $\Delta_0$, $\mathcal{C}$ is simply $\pm 2$ times the number of occupied bands per Fermi surface at a specific value $\mu$. For $\mu<\gamma_1$ we have one occupied band in ABC-stacked graphene and two bands for ABA stacking (see Appendix \ref{appa}). For ABC-stacking the band is quasi-flat close to the two Dirac points, while for ABA-stacking  we have a linear band and a quadratic band. Each band exhibits two Fermi surfaces close to the two $K$ and $K^\prime$ Dirac points and is also spin degenerate. Hence the value $\mathcal{C}=4$ for ABC-stacked graphene, and $\mathcal{C}=8$  for ABA-stacked graphene. For the yellow region $\mu\lesssim t$ we have three filled bands and thus $\mathcal{C}=12$ for all types of trilayers. For the stairway regions around $\mu= t$ there can either be two filled bands with two Fermi surfaces each centered at the $K$ and $K^\prime$ points respectively and one band with a single Fermi surface centered around $\Gamma$, or vice-versa, giving  $\mathcal{C}=2 (2\times 2-1) = 6$ and $\mathcal{C}=2(2-2)=0$ respectively. For large $\mu \gtrsim t$ we always have three bands with their Fermi surfaces centered around $\Gamma$ yielding $\mathcal{C}=-6$. This means that the two dominating regions in the phase diagram have Chern numbers either $4$ times (yellow) or $-2$ times (purple) the number of layers. This result also hold for tetra- and pentalayer graphene, see Appendix \ref{appc}.

\begin{figure}[t]
%{\sc\footnotesize \hspace*{0.2cm}ABA-stacked\hspace{2.6cm}ABC-stacked}
\begin{tikzpicture}
\node at (0,0) {
\includegraphics[height=3.5cm]{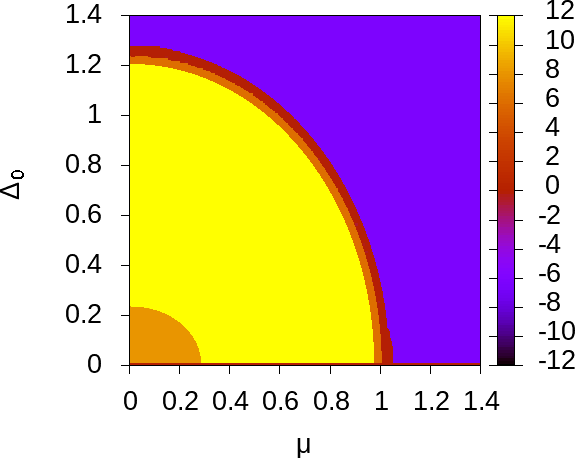}
\includegraphics[height=3.5cm]{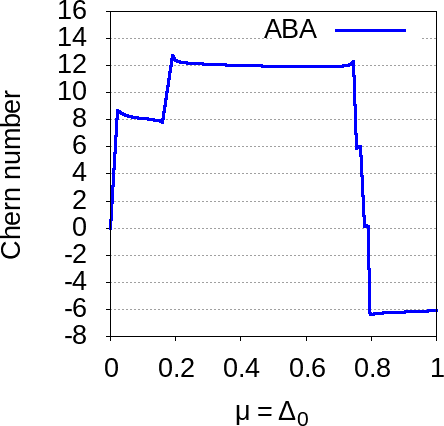}
};
\node at (-4,1.5) {(a)};
\end{tikzpicture}
\begin{tikzpicture}
\node at (0,0) {
\includegraphics[height=3.5cm]{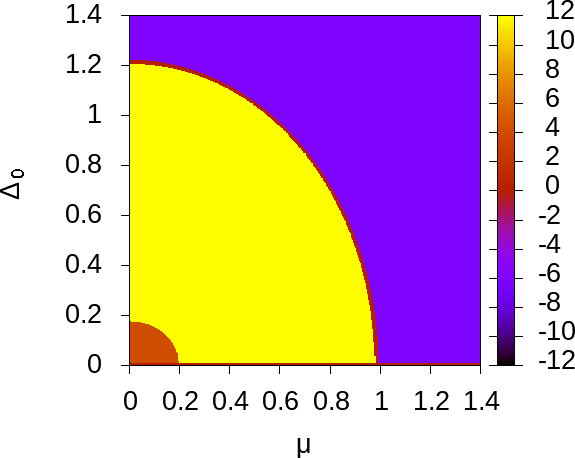}
\includegraphics[height=3.5cm]{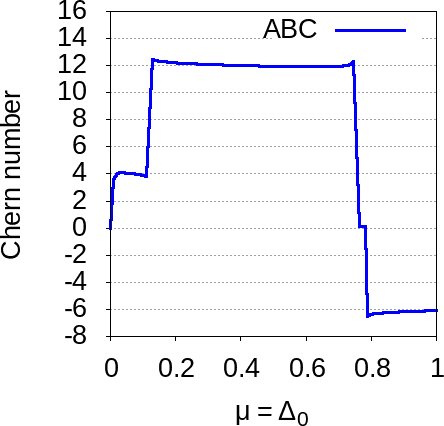}
};
\node at (-4,1.5) {(b)};
\end{tikzpicture}
\caption{(Left) Chern number $\mathcal{C}$ for (a)~ABA-stacked and (b)~ABC-stacked trilayer graphene  with $p+ip'$ SC order parameter as a function of $\mu$ and $\Delta_0$ and (Right) Chern number along the diagonal $\mu=\Delta_0$ for $\gamma_1=0.2$ and $\gamma_3=0.2$.}\label{figure_chern_trilayer_warping}
\end{figure}
For trilayer graphene with warping included, here using $\gamma_3=0.2$,  the dark-orange and red regions separating the yellow region with $\mathcal{C} = 12$ and purple region  with $\mathcal{C} = -6$ are strongly modified as seen in Fig.~\ref{figure_chern_trilayer_warping}. In the case of ABA-stacked trilayer, the widths of the dark-orange and red regions are both reduced, while for ABC-stacked trilayer the width of the red region is reduced whereas the dark-orange region entirely disappears. For completeness we give the critical parameters $\Delta_c$ and $\mu_c$ characterizing the boundaries of equation $\Delta_0=\Delta_c\sqrt{1-(\mu/\mu_c)^2}$ between the regions with different values for $\mathcal{C}$ in in Table~\ref{table3}.

\begin{table}
\begin{tabular}{|l|c|c|}
  \hline
 &$\quad\Delta_c\quad$ & $\quad\mu_c\quad$ \\
   \hline
 Normal ABA trilayer  ($\gamma_1=0.2$ and $\gamma_3=0$) & 
 \begin{tabular}{c}
 0.23\\ 1.21\\ 1.24 \\ 1.3
\end{tabular}
 & 
  \begin{tabular}{c}
0.28\\ 0.86\\ 1 \\1.15
\end{tabular}\\
 \hline
  Normal ABC trilayer  ($\gamma_1=0.2$ and $\gamma_3=0$) & 
 \begin{tabular}{c}
 0.17\\ 1.21\\ 1.24 \\ 1.3
\end{tabular}
 & 
  \begin{tabular}{c}
0.19\\ 0.86\\ 1 \\1.15
\end{tabular}\\
 \hline
Warped ABA trilayer  ($\gamma_1=\gamma_3=0.2$) & 
 \begin{tabular}{c}
0.23\\ 1.2\\ 1.23\\ 1.27
\end{tabular}
&
 \begin{tabular}{c}
 0.28\\ 0.97\\ 1\\ 1.03
\end{tabular}
\\
  \hline
  Warped ABC trilayer  ($\gamma_1=\gamma_3=0.2$) & 
 \begin{tabular}{c}
 0.17\\ 1.2\\ 1.23
\end{tabular}
&
 \begin{tabular}{c}
 0.19\\ 0.97\\ 1
\end{tabular}
\\
  \hline
\end{tabular}\caption{Values of the critical parameters $\Delta_c$ and $\mu_c$ for ABA- and ABC-stacked trilayer graphene for $p+ip'$ pairing.}
\label{table3}
\end{table}

\subsection{Trilayer graphene edge states}

\begin{figure}[h]
{\sc\footnotesize \hspace*{0.7cm}ABA-stacked\hspace{2.4cm}ABC-stacked}
\begin{tikzpicture}
\node at (0,0) {
\includegraphics[width=4.2cm]{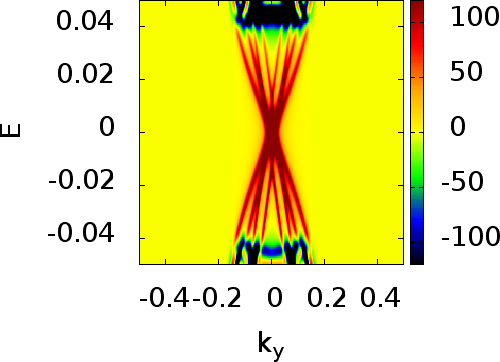}
\includegraphics[width=4.2cm]{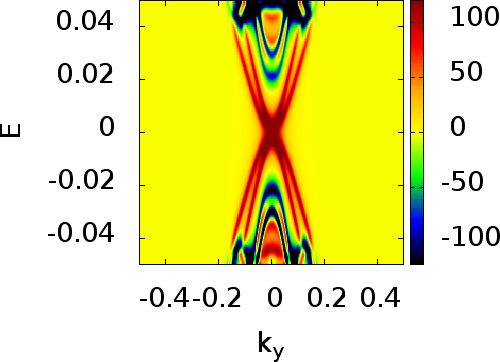}
};
\node at (-4.1,1.5) {(a)};
\end{tikzpicture}
\begin{tikzpicture}
\node at (0,0) {
\includegraphics[width=4.2cm]{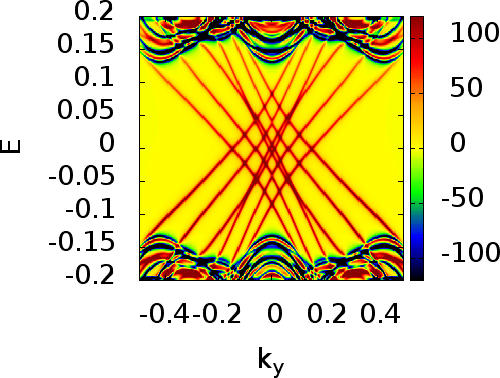}
\includegraphics[width=4.2cm]{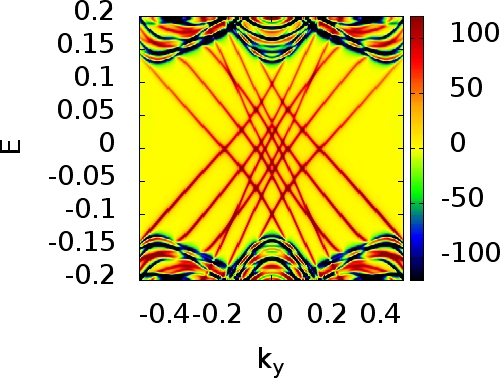}
};
\node at (-4.1,1.5) {(b)};
\end{tikzpicture}
\begin{tikzpicture}
\node at (0,0) {
\includegraphics[width=4.2cm]{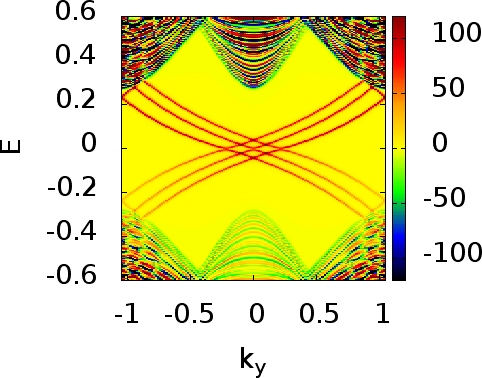}
\includegraphics[width=4.2cm]{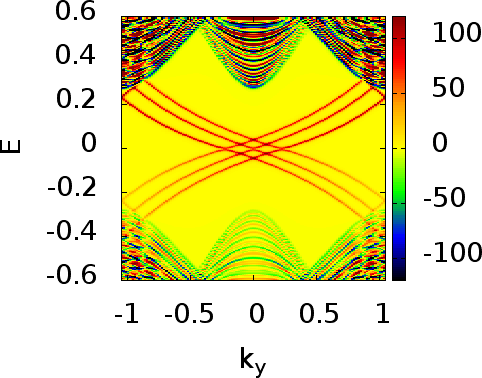}
};
\node at (-4.1,1.5) {(c)};
\end{tikzpicture}
\caption{Edge correction to the spectral function for trilayer graphene with a $p+ip'$-wave SC order parameter in trilayer graphene with armchair edges with $\gamma_1=0.2$ and $\gamma_3=0$ at (a) $\Delta_0=\mu=0.1$ corresponding to $\mathcal{C}=8$ for ABA-stacked and to $\mathcal{C}=4$ for ABC-stacked trilayer, (b) $\Delta_0=\mu=0.4$  corresponding to $\mathcal{C}=12$, and (c) $\Delta_0=\mu=1.1$ corresponding to $\mathcal{C}=-6$.}
\label{figure_p+ip'_trilayer_without_warping}
\end{figure}

Finally we also report the edge state spectrum for trilayer graphene with a NN $p+ip'$-wave SC order parameter.
The correction to the spectral function for the trilayer graphene with a line impurity along an armchair edge is plotted in Fig.~\ref{figure_p+ip'_trilayer_without_warping} 
%in the absence of warping and in Fig.~\ref{figure_p+ip'_trilayer_with_warping} in the presence of warping 
for both ABA-stacked and ABC-stacked stacking. For $\mu=\Delta_0=0.1$ (a), we obtain 8 edge states for ABA-stacking, whereas we find 4 edge state for ABC-stacking, in full agreement with the Chern number $\mathcal{C}$ extracted from Fig.~\ref{figure_chern_trilayer}. For $\mu=\Delta_0=0.4$ and $\mu=\Delta_0=1.1$, both ABA-stacked and ABC-stacked trilayer graphene exhibit 12 and 6 edges states, respectively, again in perfect agreement with the value of $\mathcal{C}$ calculated in the phase diagram in Fig.~\ref{figure_chern_trilayer}. 
%The presence of warping leads to a distortion of the edge states compared to the ones obtained in the absence of warping, but no substantial other changes for these reported values of $\mu$ and $\Delta_0$. 
Similar to the cases of monolayer and bilayer graphene, we observe also for trilayer graphene a gap closing each time that the number of edge states or Chern number changes.

\section{Summary}

We calculated the Chern number and the edge state spectral function for monolayer, bilayer and ABA- and ABC-stacked trilayer graphene in the presence of a chiral $p+ip'$- or $d+id'$-wave SC order parameter. We showed that the absolute value of the Chern number is in full agreement with the number of edge states. We  also showed that for the $p+ip'$-wave state one or more phase transitions occur in the Chern number when turning the SC order parameter amplitude $\Delta_0$ or the chemical potential $\mu$, with the transitions corresponding to gap closings in the energy spectrum. For multilayer graphene with $p+ip'$-wave pairing, the number of regions in the phase diagram increases with the number of layers, with the size of the regions  strongly affected by the trigonal warping. Notably, the largest phases have a Chern number either four times or minus twice the number of layers. In addition, we noted the existence of a peculiar low-energy region whose Chern number ($4$) does not depend on the number of layers for ABC-stacking and whose size is proportional to the value of interlayer coupling. Furthermore, we find a strong difference between NN and NNN $p+ip'$-wave pairing. This difference stems mainly from the fact that for NNN coupling the phases, as well as the gap closings \cite{Pangburn2022}, are not affected by the value of the SC order parameter amplitude, but only by the value of the chemical potential. We also took into account a phase difference of $\pi$ between layers in bilayer systems and and then found that this suppresses the topological character of both the $p+ip'$- and the $d+id'$-wave SC systems at small values of the order parameter amplitude. Last but not least we showed that the topological properties of all graphene systems at not too large SC gap values, in particular the Chern number and the number of edge states, can be determined directly from the topology of the normal state Fermi surface.

\acknowledgments We would like to thank Nathan Campion for his contribution to the parallelization of our codes. We  acknowledge financial support from the Swedish Research Council (Vetenskapsr\aa det Grant No.~2018-03488) and the Knut and Alice Wallenberg Foundation through the Wallenberg Academy Fellows program and thank the Centre de Calcul Intensif d'Aix-Marseille for granting access to high performance computing resources. NS would like to thank the National Science Centre (NCN, Poland) for funding under the grant 2018/29/B/ST3/01892.

%%%%%%%%%%%%%%%%%%%%%%%%%%%%%%%%%%%%%%%%%%%%%%%%%%%%%%%%%%%%%%%%%%%%%%%%%%%%%%%%%%%%%%%%%%%%%%%%%%%%%%%%%%%%
%                                                                                                          %
%                                                                                                          %
%                           APPENDICES                                                         %
%                                                                                                          %
%                                                                                                          %
%%%%%%%%%%%%%%%%%%%%%%%%%%%%%%%%%%%%%%%%%%%%%%%%%%%%%%%%%%%%%%%%%%%%%%%%%%%%%%%%%%%%%%%%%%%%%%%%%%%%%%%%%%%%

\begin{appendix}

\section{Normal state band structure}\label{appa}

\begin{figure}[h!]
 \includegraphics[width=5cm]{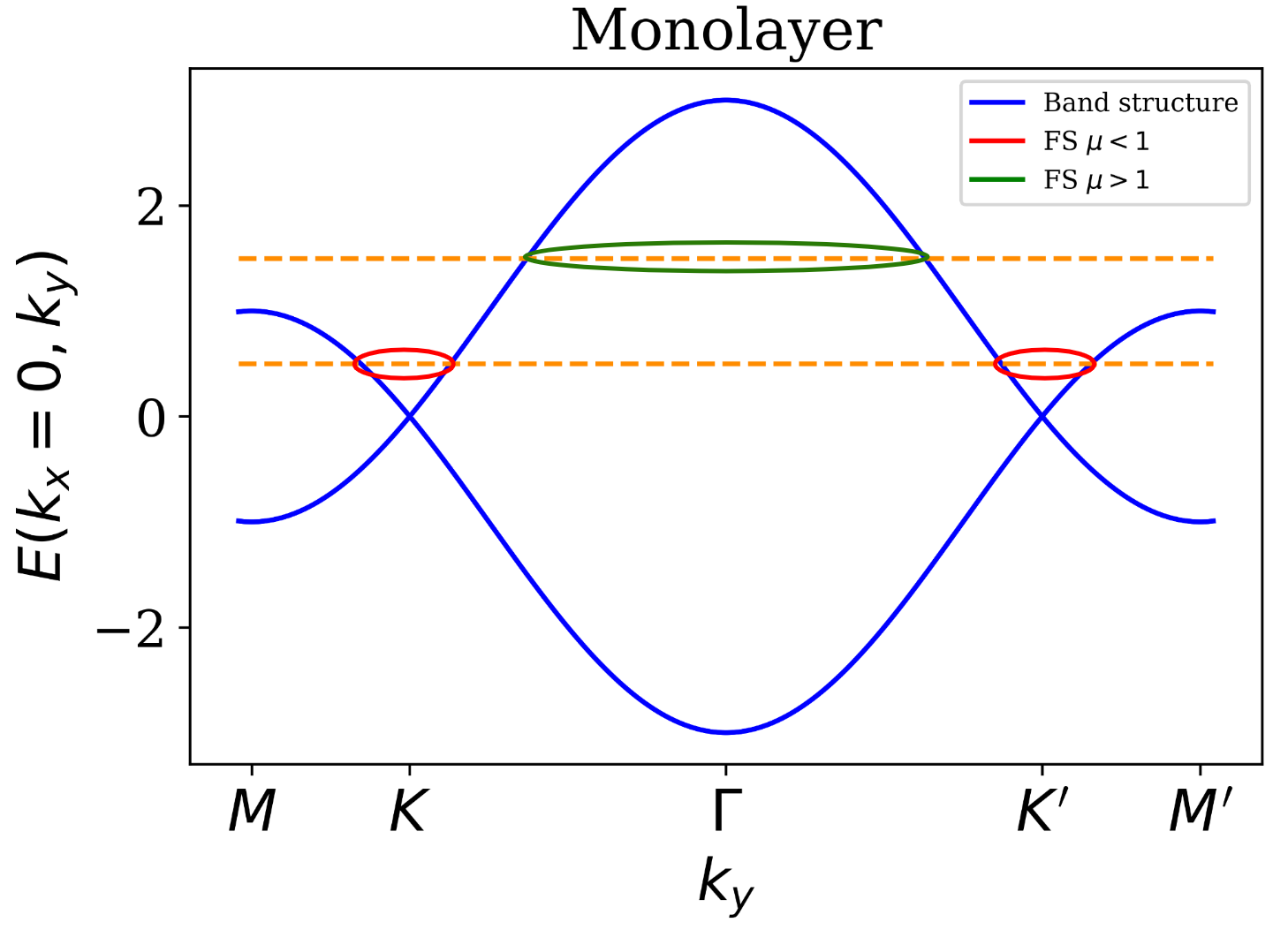}
 \caption{(a) Band structure for monolayer graphene. The orange dotted lines denote the Fermi level for a regime in which the normal state Fermi surface is described by two circles centered around the $K$ points (denoted in red, $\mu<1$), and respectively a single circle centered around the $\Gamma$ point ($\mu>1$).
 \label{monolayer_ns}}
\end{figure}

\begin{figure}[h!]
 \includegraphics[width=4cm]{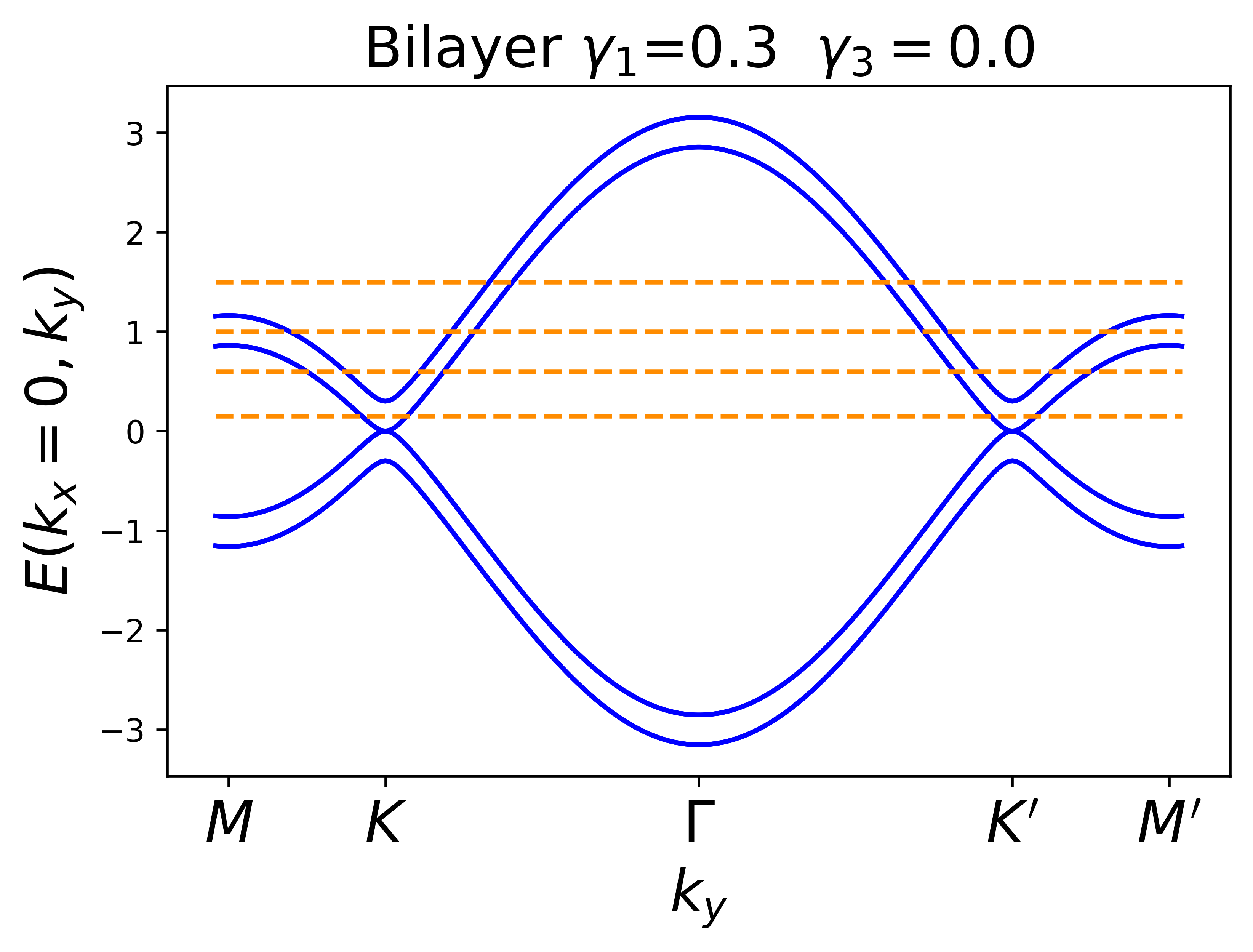}
 \includegraphics[width=4cm]{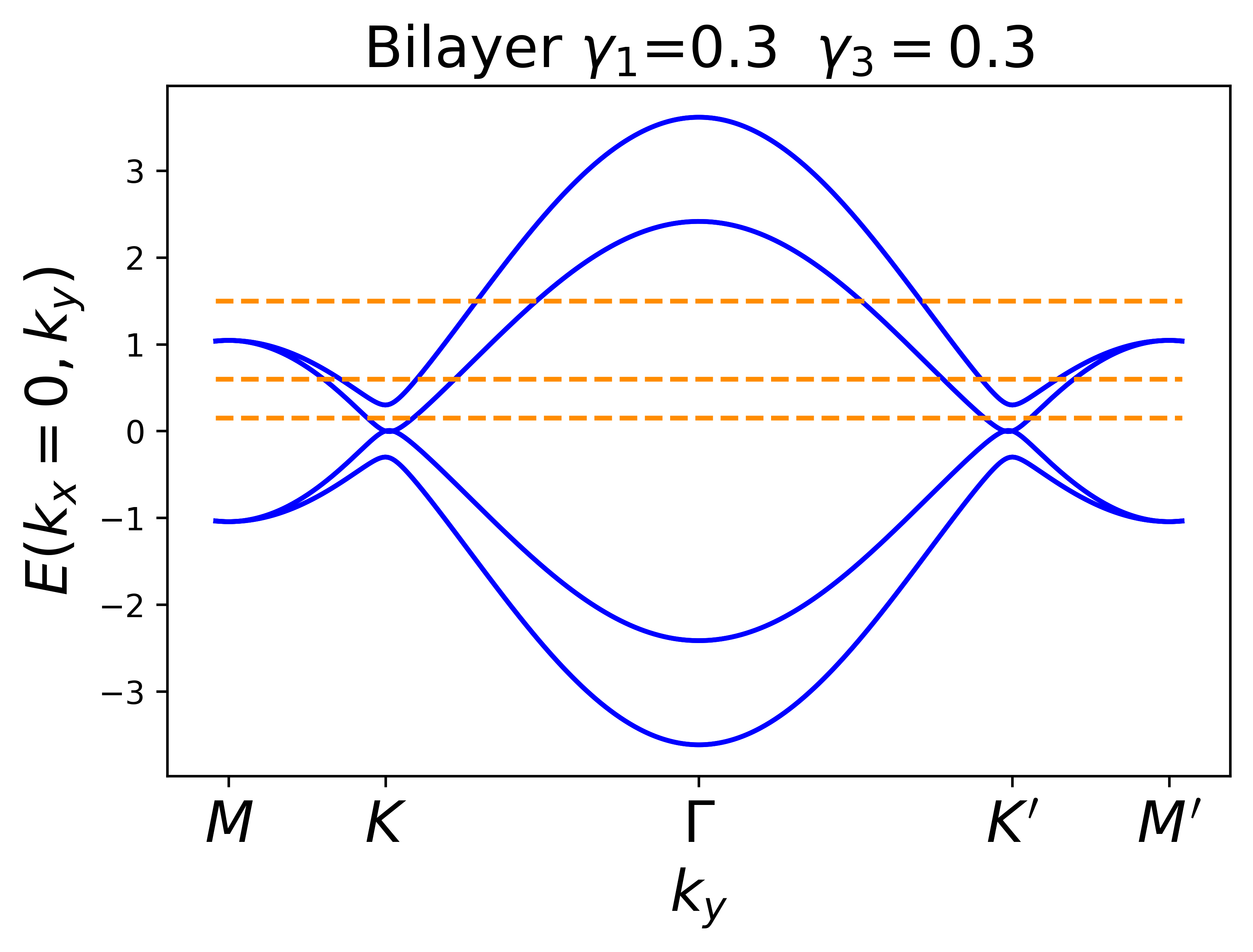}
 \caption{Band structure for bilayer graphene (left) without trigonal warping, and (right) with trigonal warping. Note that close to the M point the splitting of the bands is reduced in the presence of $\gamma_3$, consistent with the reduction of the size of the regions with a Chern number of 6 and 0.}
 \label{bilayer_ns_tw}
\end{figure}

\begin{figure}[h!]
 \includegraphics[width=4cm]{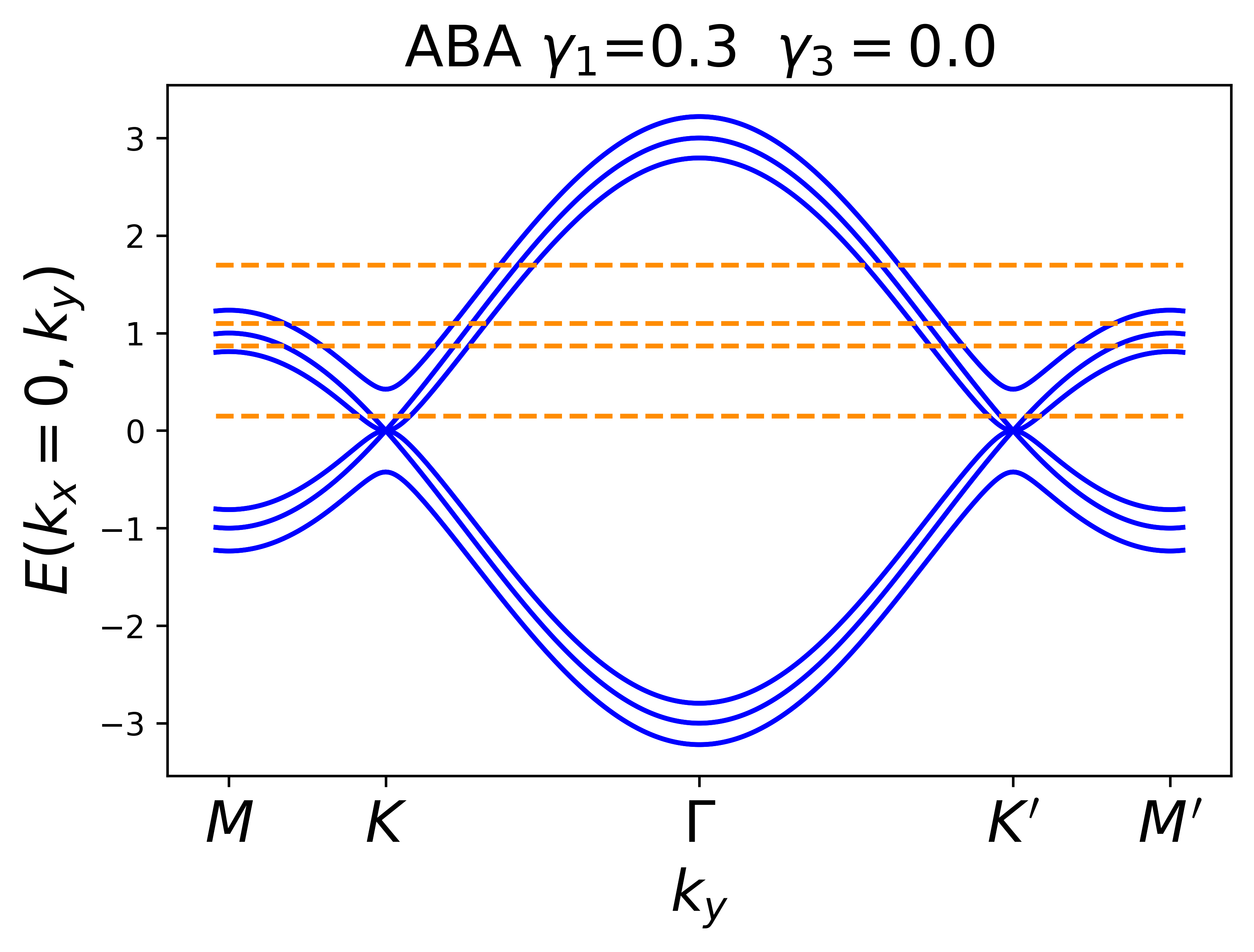}
 \includegraphics[width=4cm]{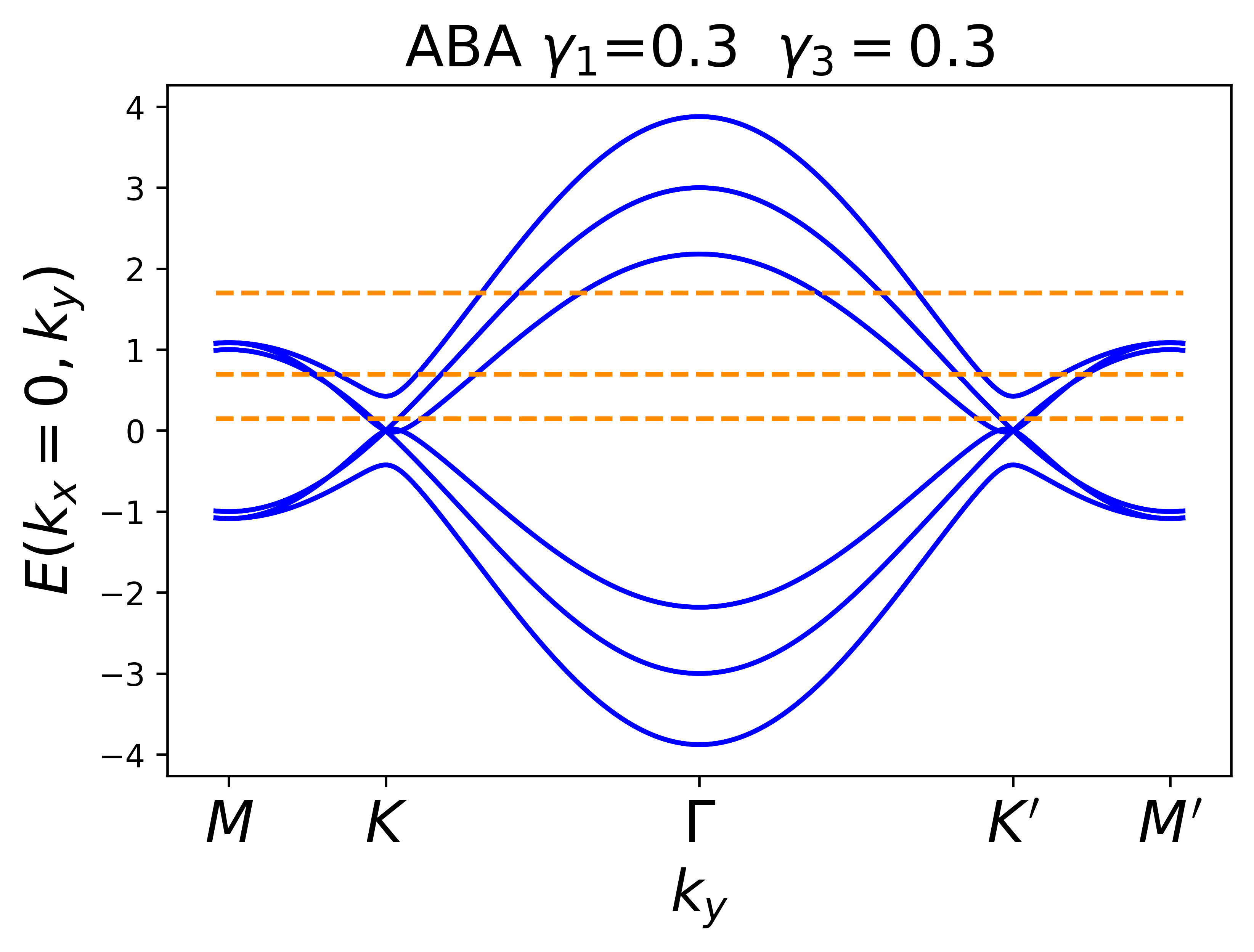}
 \caption{Band structure for ABA trilayer graphene (left) without trigonal warping, and (right) with trigonal warping. }
 \label{ABA_ns_tw}
\end{figure}

\begin{figure}[h!]
 \includegraphics[width=4cm]{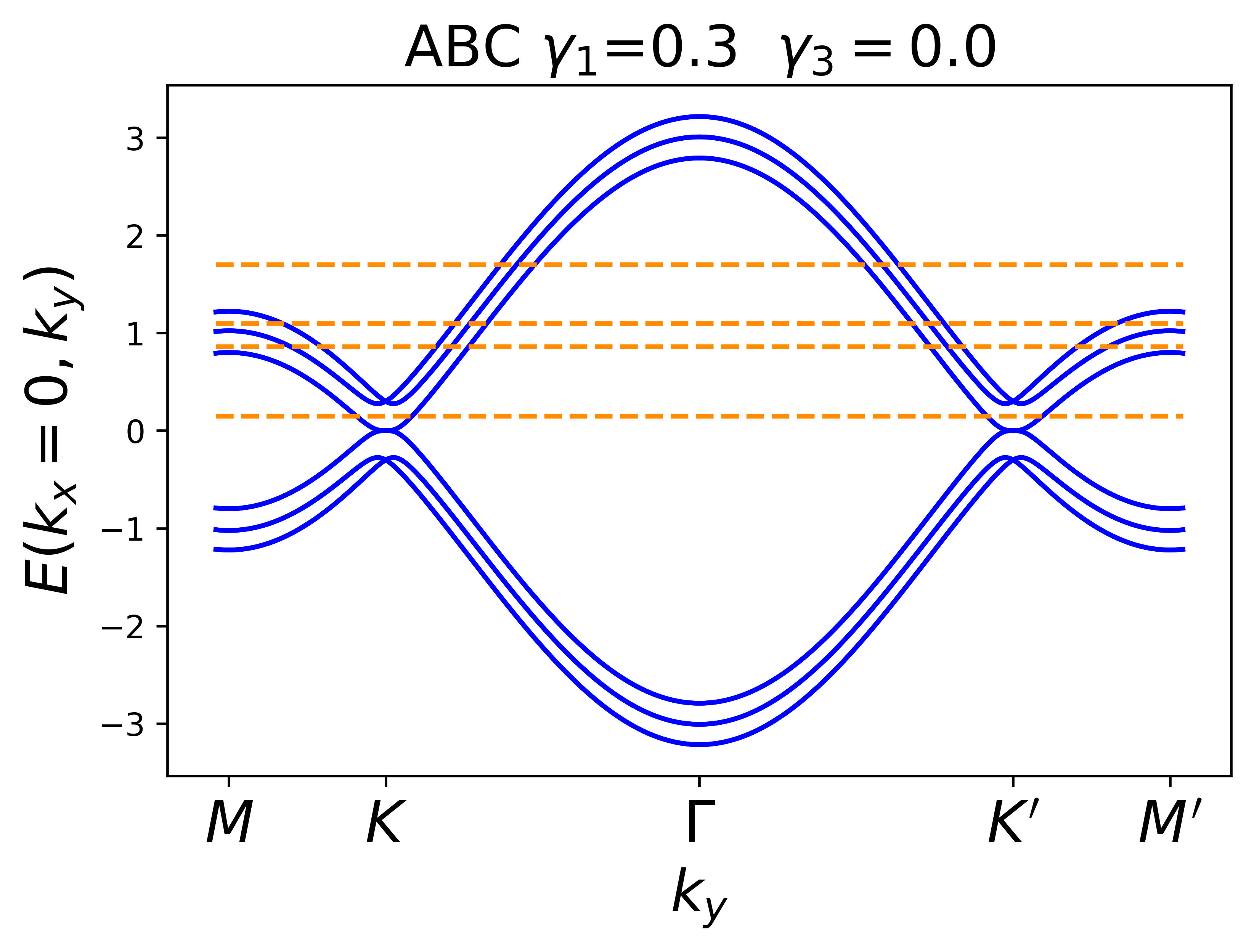}
 \includegraphics[width=4cm]{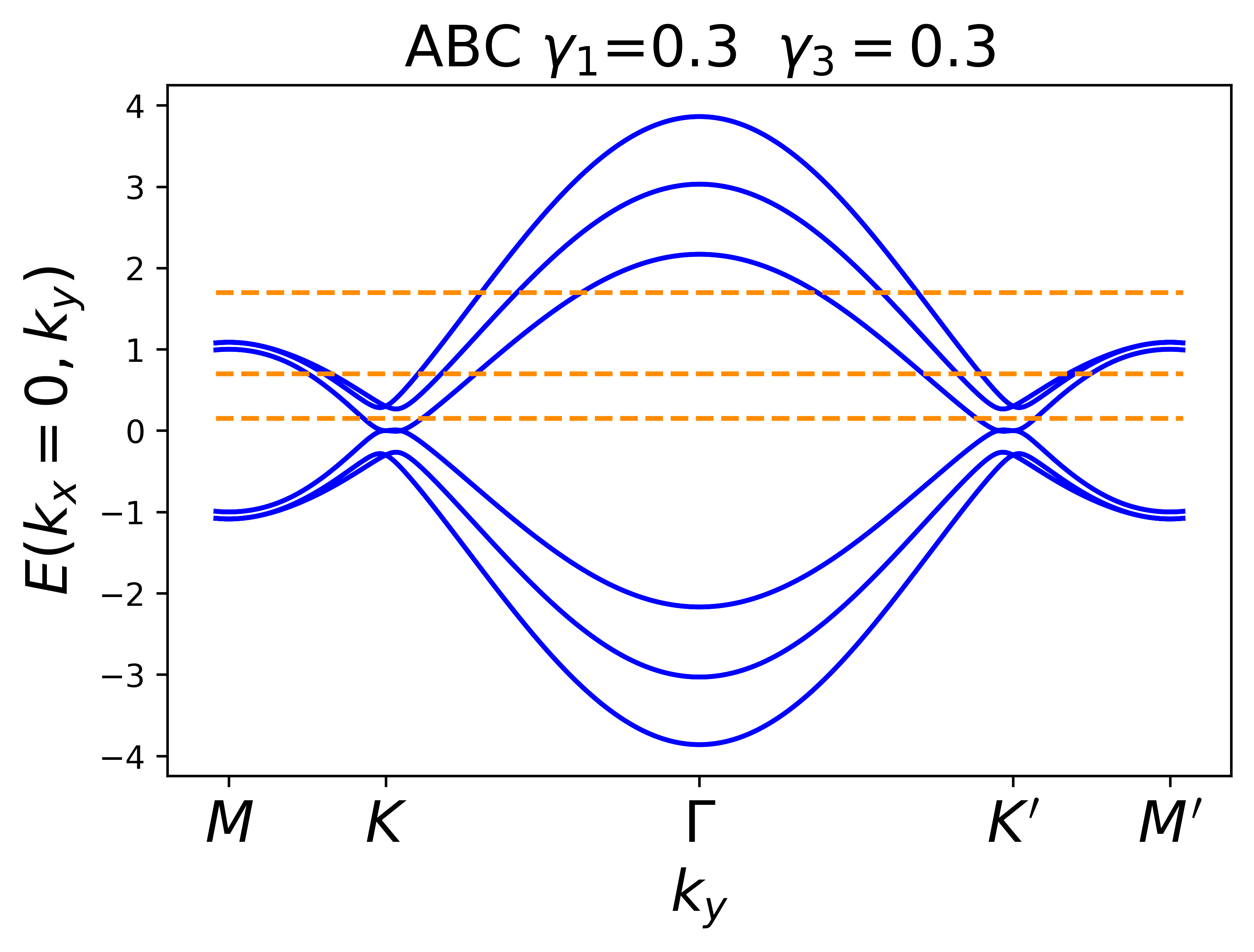}
 \caption{Band structure for ABC trilayer graphene (left) without trigonal warping, and (right) with trigonal warping}
 \label{ABC_ns_tw}
\end{figure}

Figures~\ref{monolayer_ns} to \ref{ABC_ns_tw} give the normal state band structure for monolayer graphene, as well as for bilayer graphene, and trilayer graphene with ABA and ABC stacking. This will help us to make the correspondence to the Chern number in the $p+ip'$ SC state, as indicated in the main text. The orange dotted lines denote the Fermi level for the various regimes described in the main text, indicating the number of filled bands and respectively the topology of the normal state Fermi surface, for example if it corresponds to one or two separate regions.

\section{Chern number for bilayer and trilayer graphene with a $d+id'$-wave SC order parameter}\label{appb}

\begin{figure}[h]
%{\sc\footnotesize \hspace*{0.2cm}ABA-stacked\hspace{2.6cm}ABC-stacked}
\begin{tikzpicture}
\node at (0,0) {
\includegraphics[width=4cm]{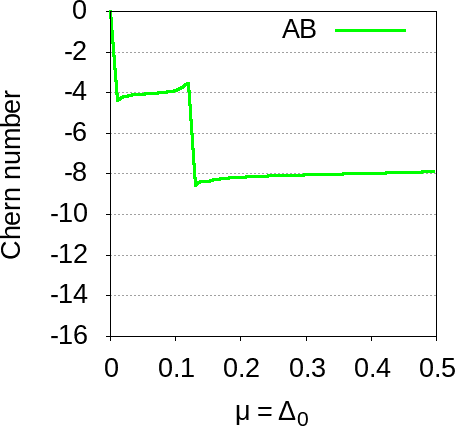}
%\hspace{0.5cm}
\includegraphics[width=4cm]{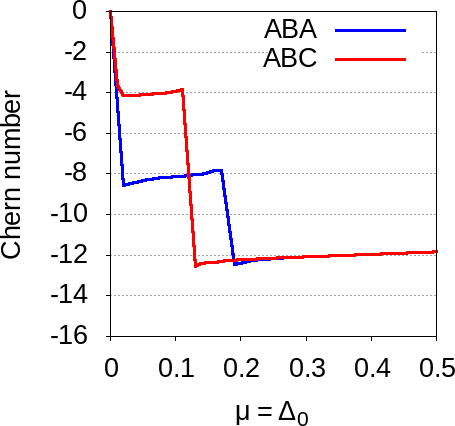}
};
\node at (-3.8,1.7) {(a)};
\node at (0.3,1.7) {(b)};
\end{tikzpicture}
\caption{Chern number $\mathcal{C}$ for (a) bi- and (b) tri-layer graphene with a $d+id'$-wave SC order parameter, plotted along the diagonal $\mu=\Delta_0$ at $\gamma_1=0.2$ and $\gamma_3=0$.}\label{fig_appb}
\end{figure}

For bilayer and trilayer graphene with a $d+id'$-wave SC order parameter, the Chern number $\mathcal{C}$ varies with $\mu$ and $\Delta_0$, contrary to what one has found for monolayer graphene, for which $\mathcal{C}$ is constant and equal to $-4$, see Fig.~\ref{figure_chern_monolayer}(a). Indeed, for bilayer graphene in the low ($\mu,\Delta_0$) region, $\mathcal{C}$ is equal to $-4$, and it changes to $-8$ with increasing $\mu=\Delta_0$, see Fig.~\ref{fig_appb}(a). For trilayer graphene in the low ($\mu,\Delta_0$) region, $\mathcal{C}$ is equal to $-4$ or $-8$ for an ABC, and respectively an ABA stacking. It subsequently changes to $-12$ for large $\mu=\Delta_0$, see Fig.~\ref{fig_appb}(b).

\section{Chern number for tetra- and pentalayer graphene with $p+ip'$-wave SC order parameter}\label{appc}

\begin{figure}[H]
%{\sc\footnotesize \hspace*{0.2cm}ABA-stacked\hspace{2.6cm}ABC-stacked}
\begin{tikzpicture}
\node at (0,0) {
\includegraphics[width=4cm]{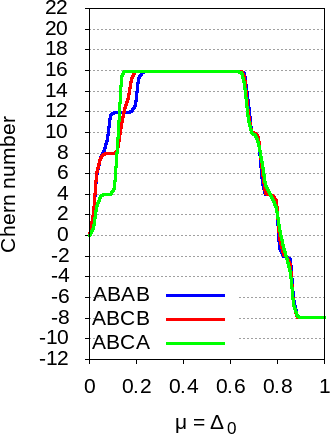}
%\hspace{0.5cm}
\includegraphics[width=4cm]{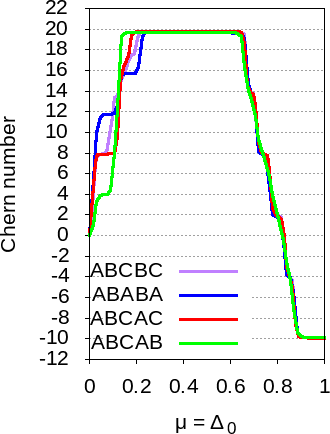}
};
\node at (-3.8,2.4) {(a)};
\node at (0.3,2.4) {(b)};
\end{tikzpicture}
\caption{Chern number $\mathcal{C}$ for (a) tetra- and (b) penta-layer graphene with a $p+ip'$-wave SC order parameter, plotted along the diagonal $\mu=\Delta_0$ at $\gamma_1=0.2$ and $\gamma_3=0$ with the possible stacking choices\cite{Wirth2022} in different colors.}\label{fig_C_multilayers}
\end{figure}

In this Appendix we present the Chern number $\mathcal{C}$, extracted at $\mu=\Delta_0$ for both tetralayer and pentalayer graphene with a $p+ip'$-wave SC order parameter. We plot the Chern number  in Fig.~\ref{fig_C_multilayers}, with the results summarized in Tab.~\ref{table_low_energy}. 
We conclude that there are only a finite number of possible values for the Chern number in region (i), i.e the region with $\mu, \Delta_0 \lesssim \gamma_1$: $\mathcal{C}=4 n$ with $n \in[1,L-1]$ being the number of occupied bands with two Fermi surfaces centered around the $K$ and $K^\prime$ points, and $L$ the number of layers. Note that a Chern number of~$4$ is only possible for a fully rhombohedral stacking, i.e.~ABCA for tetralayer and ABCAB for pentalayer graphene. For rhombohedral graphene this value does not depend on the number of layers and is unique in region~(i). On the other hand the Chern number in the intermediate $\mu$/$\Delta_0$ region~(ii), for which $\gamma_1\lesssim\mu,\Delta_0 \lesssim t$, is given by four times the number of layers: $\mathcal{C}=4L$ because all the filled bands for all types of stacking have two Fermi surfaces centered around the two $K$ and $K^\prime$ points and a Chern number of $4$ each. In the large  $\mu$/$\Delta_0$ region~(iii), $\mu,\Delta_0 \gtrsim t$, the Chern number seems also to follow a similar pattern being given by two times of number of layers: $\mathcal{C}=-2L$ because all occupied bands have a Fermi surface centered around the $\Gamma$ point and a Chern number of $-2$. The stairway region in-between regions (ii) and  (iii) situated in the vicinity of $\mu, \Delta \approx t$, show a quantized value of $\mathcal{C}$ equal to $4(L-\ell)-2\ell=4L-6\ell$ with $\ell\in[1,L-1]$ being the number of filled bands with one Fermi surface centered around the $\Gamma$ point for which $ \mathcal{C}=-2$, and $L-\ell$ being the number of filled bands with two Fermi surfaces centered around the $K$ and $K^\prime$ points for which $ \mathcal{C}=4$.

\begin{table}
\begin{tabular}{|l|c|c|c|}
\hline
System &  Minimal value (i) &Region (ii) & Region (iii) \\
\hline
  \hline
  Monolayer & 4 & 4 & $-2$ \\
\hline
  \hline
  Bilayer AB & 4 & 8 & $-4$ \\
\hline
  \hline
  Trilayer ABA & 8 & 12 & $-6$ \\
     \hline
  Trilayer ABC & 4 & 12 & $-6$   \\
\hline
  \hline
  Tetralayer ABAB & 12 & 16 & $-8$   \\
        \hline
  Tetralayer ABCB & 8 & 16 & $-8$   \\
       \hline
 Tetralayer ABCA & 4 & 16 & $-8$  \\
\hline
  \hline
Pentalayer ABCBC & 8 & 20 & $-10$ \\
        \hline
  Pentalayer ABABA & 12 & 20 & $-10$ \\
       \hline
 Pentalayer ABCAC & 8 & 20 & $-10$  \\
       \hline
Pentalayer ABCAB & 4 & 20 & $-10$  \\
       \hline
\end{tabular}
\caption{Chern number $\mathcal{C}$ for monolayer and different multilayer graphene systems in the three largest regimes of the ($\mu$,$\Delta_0$) phase diagram: the minimal Chern value in the low $\mu$ region (i) with $\mu,\Delta_0\lesssim\gamma_1$, the Chern value in the intermediate region (ii) with $\gamma_1\lesssim\mu,\Delta_0 \lesssim t$, and the Chern value in the high $\mu$ region (iii) with $\mu,\Delta_0 \gtrsim t$.}
\label{table_low_energy}
\end{table}
\end{appendix}

%\section{Flat band enhancement by a displacement field}\
%\label{appb}
%\textcolor{red}{Don't see this in the main text, so skip entirely here?}

%%%%%%%%%%%%%%%%%%%%%%%%%%%%%%%%%%%%%%%%%%%%%%%%%%%%%%%%%%%%%%%%%%%%%%%%%%%%%%%%%%%%%%%%%%%%%%%%%%%%%%%%%%%%
%                                                                                                          %
%                                                                                                          %
%                           REFERENCES                                                       %
%                                                                                                          %
%                                                                                                          %
%%%%%%%%%%%%%%%%%%%%%%%%%%%%%%%%%%%%%%%%%%%%%%%%%%%%%%%%%%%%%%%%%%%%%%%%%%%%%%%%%%%%%%%%%%%%%%%%%%%%%%%%%%%%

\bibliographystyle{apsrev4-1}
%\bibnote[]{} pour inserer une note dans le texte.

\end{document}